# Electrically activated W-doped VO$_2$ films for reliable, large-area, broadband THz waves modulators

*Eduard-Nicolae Sirjita[a,b]\*, Alexandre Boulle[b]\*, Jean-Christophe Orlianges[a], Richard Mayet[b], Aurélien Debelle[c], Lionel Thomé[c], Maggy Colas[b], Julie Cornette[b], Aurelian Crunteanu[a]\**

[a] XLIM Research Institute, CNRS/ University of Limoges, France;

[b] Institut de Recherche sur les Céramiques (IRCER), CNRS UMR 7315, University of Limoges, France

[c] IJCLab, Université Paris-Saclay, CNRS/IN2P3, Orsay, France.

**Abstract**

THz amplitude modulators and switches are considered to be the main building blocks of future THz communication systems. Despite rapid progress, modulation and switching devices in this electromagnetic spectrum lag far behind other frequency ranges. Currently, THz modulators face major challenges in consistently producing high modulations depths over large frequency bands. Moreover, a convenient integration for practical applications requires that the modulation/switching properties can be electrically controlled. Devices fulfilling all these conditions remain to be demonstrated. In this work we show that W-doped VO$_2$ films grown by direct-current magnetron sputtering can be efficiently used for the development reliable, large-area, broadband THz waves modulators. We demonstrate that W doping not only permits to tune the insulator to metal transition (IMT) temperature of $VO_2$, but also, most importantly, to control the topology of the electrically activated transition. In situ / operando X-ray diffraction and Raman spectroscopy characterizations of the devices, coupled with standard resistivity measurements and time-domain THz spectroscopy, unambiguously demonstrate that the changes in the spatial distribution of the IMT is due to structural distortions induced by W doping. These findings are exploited to validate $VO_2$-based devices whose IMT that can be triggered either thermally or electrically over areas as large as 3.8×10 mm$^2$, hence permitting the development of efficient THz modulators operating over a large spectral range (0.2 – 2 THz) with MDs reaching 96%.

1. **Introduction**

The field of terahertz (THz) science and technology is constantly attracting increasing attention as the electromagnetic radiation in this frequency range (1-10 THz) has emerging applications in security, short-range telecommunications, medical applications, and quality control of food and materials[1,2]. Despite rapid progress, modulation and switching devices in this electromagnetic spectrum lag behind other frequency ranges[3]. THz amplitude modulators are considered to be the building blocks of future THz systems, especially in communication systems. The main candidates employed for THz modulation are based on semiconductors (Si, AlGaAs-…), 2D materials (graphene, $MoTe_2$…) heterostructures, metamaterial-based structures, and phase-change materials, such as vanadium dioxide ($VO_2$), as attested in recent literature reviews [3–5]. The key factors that are considered to be of importance for amplitude modulation are the modulation depth (MD), the operation frequency domain, the reconfiguration speed, and the power consumption.

Currently, THz modulators still face challenges in consistently producing high MDs over large frequency bands and often require relatively expensive fabrication techniques. Furthermore, large MDs are often produced by optically irradiating the substrate material in order to generate carriers that affect the conductivity of the thin film, thus changing the THz modulation performances[5]. Introducing an external light source is, however, inherently increasing the complexity of the system. A material that stands out for THz modulation is $VO_2$[6–8] due to its various advantages such as the abrupt change between its insulator and metallic states using optical, electrical or thermal stimuli, its high energy efficiency[3] and its ability to operate on different broadband frequency domains. These performances make the integration of $VO_2$ highly attractive for a variety of applications spanning from electrical memory devices to thermochromics, thermal regulation and reconfigurable high frequency devices[9,10].

The physical property of $VO_2$ of interest for THz modulation is its reversible and ultrafast transition between an insulating and a metallic phase that is triggered at a relatively low transition temperature, $T_{IMT}$, of 340K[11]. The insulator to metal transition (IMT) is also accompanied by a structural phase transition (SPT) where $VO_2$ changes from a monoclinic M1 structure at low temperatures to a tetragonal rutile-like R structure at temperatures higher than the transition temperature. The IMT/ SPT can be triggered by direct heating above the transition temperature or by applying a sufficiently high electrical stimulus[12]. When cooling down the material from the metallic phase to the initial insulating phase, a slight offset in the reverse metal-to-insulator temperature is observed, creating a well-known hysteresis effect. As the name suggests, the key intrinsic property that changes with the IMT is the electrical resistivity, with a ratio of more than five orders of magnitude between the two states[13]. From the optical properties point of view, $VO_2$ changes from a high to a low optical transmission in the visible and infrared domains[14,15]. Besides the previously mentioned applications, the IMT of $VO_2$ has been successfully exploited for THz modulation devices with bare films' MDs going as high as ~85%[16,17], which can be further increased to values of ≥ 90% when incorporated in resonant-type metamaterials designs, but at the expense of the operation bandwidth reduction[8,16].

One particular appealing property of $VO_2$ is the ability to tune the parameters that define the IMT such as the transition temperature ($T_{IMT}$), the hysteresis width ($\Delta H$) or the transition width ($\Delta T$), in order to customize $VO_2$ for different applications. This can be achieved in several ways, for instance by controlling the oxygen stoichiometry[18], by strain engineering[19] or metal doping[20] or by adjusting the experimental parameters (ex. deposition and annealing temperatures) during the film deposition process[13]. However, the most efficient method of achieving a drastic modification of the hysteresis curve characteristics is the incorporation of metallic dopants in the structure of the vanadium dioxide film. As such, it was previously reported that dopants like Al[21], Fe[22] or Cr[23] will increase the $T_{IMT}$ while W[24], Mo[25], Mg[26] and Nb[20] dopants are likely to reduce the $T_{IMT}$. Although metal doping has proved to be an efficient way to reduce the overall energy budget associated with the IMT by drastically lowering the $T_{IMT}$ of the films it has, however, a detrimental effect for THz modulating applications, by also reducing the MD[20]. One of the most investigated dopants is tungsten, known to usually reduce the $T_{IMT}$ of $VO_2$ thin films by 20- 28 K/at. %W doping [27,28].

The influence of W atoms over the IMT modulation arises from an interplay between the electron doping that disrupts the dimerized V-V pairs of the room temperature phase and increases the conductivity in the films, and a local distortion of the monoclinic phase toward a rutile like phase[27,29,30]. This rutile like phase affects the nearby M1 lattice and reduces the potential energy barrier for phase transition[29] while also acting as a nucleation site for the R phase when the transition occurs. The exact role of electron doping is still the subject of debate; whereas some theoretical work suggest that electron doping from the W atoms have little effect on the $T_{IMT}$[28], others suggest that it is electron doping that make W a more efficient dopant at reducing the transition temperature, as compared to other atoms with large ionic radii[31].

In this work we focus on W-doped $VO_2$ epitaxial films grown on (001) oriented sapphire by direct current magnetron sputtering. We demonstrate that W doping allows not only to tune the IMT temperature, but also to control the topology of the electrically activated transition. This feature is exploited to fabricate $VO_2$ – based devices whose IMT that can be triggered either thermally or electrically over areas as large as 3.8×10 mm$^2$, hence permitting the development of efficient THz modulators operating over a large spectral range (0.2 – 2 THz) with MDs reaching 96%. Our findings are supported by *in situ* / *operando* X-ray diffraction (XRD) and Raman spectroscopy characterizations of the devices, coupled with standard resistivity measurements and time-domain THz spectroscopy.

2. **Experimental section**

*2.1. Film growth*

Tungsten-doped $VO_2$ films can be grown by various methods, including sol-gel[32], pulsed laser deposition[33] or magnetron sputtering[34]. An advantage of magnetron sputtering is that it allows to grow films, with a high degree of reproducibility, over very large areas which makes it de facto scalable to industrial scale. Reactive magnetron sputtering is a variation to the classical sputtering method where, by adding a reactive gas during the deposition process, one can incorporate, in a controlled manner, additional elements in the film during the deposition process.

In this study, $VO_2$ films were obtained using reactive DC magnetron sputtering of a vanadium target in an argon and oxygen atmosphere, on $10 \times 10$ mm$^2$ (001)- oriented $Al_2O_3$ substrates at a temperature of 500°C, with a subsequent *in situ* annealing step in pure oxygen atmosphere (20-mTorr pressure, at 550°C for 30 minutes[13]). The tungsten-doped films with three different doping levels were grown by placing metallic tungsten chips on the vanadium target, the doping level being proportional to the size of the W chips. The films are termed as D0 (undoped $VO_2$), D1, D2 and D3 for $VO_2$ films with increased W doping amounts. The average deposition rate evaluated across the different doped samples was 8.8 nm/min, with a notable exception for the D2 sample, for which we estimated a lower rate, of 4.5 nm/min, possibly due to the plasma instabilities during the sputtering process. Nonetheless, all films are thick enough to consider that the thickness-related strain has little or no influence on their hysteresis parameters [35].

With the above-mentioned deposition conditions, the VO$_2$ films exhibit a (010) orientation with respect to the (001) planes of sapphire[13]. In the in-plane direction, the epitaxial relationships are given by [100]$_{VO2}$||[210]$_{Al2O3}$, [100]$_{VO2}$||[110]$_{Al2O3}$ and [100]$_{VO2}$||[120]$_{Al2O3}$. The three orientations correspond to three different epitaxial variants which are due to the existence of three structurally equivalent orientations of the (010) planes with respect to the (001) plane of sapphire[13].

*2.2 Electrical characterizations*

The DC electrical characteristics of the vanadium oxide compositions were recorded in the 25° to 95°C temperature range using a four-point probe setup, i.e. four in-line spring-loaded probes with a spacing of 1 mm connected to a Keithley 2612B sourcemeter. A Peltier thermoelectric element was used to control the temperature during the heating and cooling steps, and a Pt100 thermocouple was used to control the temperatures.

The evolution of the resistivity of the VO$_2$ films with increasing temperature has a typical hysteresis shape (see Figure S1 in the Supporting Information). In order to quantify the parameters characterizing the transition, the curves were fitted with the following function:

$$f(x) = a \times \left[1 - tanh\left(\frac{x - b}{c}\right)\right] + d - e \times x \qquad (1)$$

where a is the amplitude of the transition, b the middle point of the transition (the inflection point of the curve), c is the width of the transition, d is a parameter that controls the vertical offset of the function and e is a parameters that controls the slope of the linear part of the transition (Figure S1 in the Supporting Information). From these, the b parameter gives directly the middle points of the heating or cooling curves (T$_{IMT}$ and T$_{MIT}$), respectively, which are also used to calculate the average transition temperature T$_{avg}$ = (T$_{IMT}$ + T$_{MIT}$)/2 and the hysteresis width, ΔH= $T_{IMT}$- $T_{MIT}$. The transition width ΔT (*i.e.*, the total temperature region where the resistivity varies non linearly) has been obtained from the second derivative of this function using a procedure described in [13]. The uncertainties on all parameters are estimated from the standard deviation of the fitted parameters.

*2.3 Structural characterizations*

The elemental composition of pure and W- doped $VO_2$ films was determined by Rutherford Backscattering Spectrometry (RBS) experiments performed at the JANNuS-SCALP facility of IJCLab in Orsay[36]. We used 1.4 MeV $He^+$ ions at normal incidence to the sample surface, and a passivated implanted planar silicon detector located at a scattering angle of 165°. The recorded RBS spectra were analysed using the SIMNRA computer code[37].

A Bruker D8 "Discover" X-ray diffractometer was used to perform X-ray diffraction (XRD) measurements. The X-ray beam from a copper target was collimated using a parabolic multi-layer mirror and a 2-reflections Ge (220) monochromator calibrated to select the $K\alpha_1$ radiation of Cu. The diffracted x-rays were collected using a 1D position sensitive detector ("Lynx Eye") with a resolution of 0.01° 2θ. An Anton Paar DHS 1100 furnace mounted to the sample stage was used to characterize the evolution of the structure of the $VO_2$ films with temperature. In the current experiment the beam size is 0.2×10 $mm^2$.

In order to quantify the relative amount of M1 and R phases in the film during the transition, the 020 peak of $VO_2(M1)$ and the 200 peak of $VO_2(R)$ have been fitted with pseudo-Voigt functions[38] to extract the integrated intensity of each phase. Since the diffracted intensity is proportional to the volume of diffracting material, the volume fraction of each phase can be computed using [35]:

$$x_{M1} = \frac{I_{M1}\, v_{M1}\, /\, |F_{020}(M1)|^2}{I_{M1}\, v_{M1}\, /\, |F_{020}(M1)|^2\, +\, I_R\, v_R\, /\, |F_{200}(R)|^2} \quad (2)$$

where $x_i$ and $v_i$ (i = $M_1$, R) are the volume fractions and unit-cell volumes of i = $M_1$ or R, and $F_{hkl}$ is the structure factor of the {hkl} lattice plane family. From the temperature evolution of the volume fraction the SPT can be quantitatively monitored and the characteristic of the transition can be estimated by fitting the data with equation (1) above. In the case of the highest doping levels, the large overlap of the $020_{M1}$ peak and and the $200_R$ peaks during the SPT makes the fitting procedure unstable. Hence, to avoid the occurrence of unphysical solutions, the positions of the $020_{M1}$ and $200_R$ peak have been constrained to their value just before and just after the transition, respectively. In the following we solely consider the M1 phase fraction as it allows for a direct comparison with the resistivity data.

The Raman spectra of the samples were acquired in the spectral range 50–1300 $cm^{-1}$ using an inVia Reflex Renishaw confocal Raman spectrometer. Raman data were collected after two accumulation spectra of 60 s each, using a ×50 LWD objective using a 532-nm excitation laser.

*2.4 THz characterizations*

The temperature- and electrical- dependent THz transmission characteristics of the layers were recorded in the 0.2 – 2 THz range using a THz time domain spectrometer (THz-TDS) TERA K15 all fiber-coupled terahertz spectrometer from Menlo Systems GmbH. A Peltier heating element with a hole in the middle, large enough to allow the transmission of the THz radiation, has been used to perform the measurement in transmission mode.

The acquired time-domain signals transmitted through the samples were transformed to the frequency domain using a Fast Fourier Transform (FFT). Afterwards, the spectra were normalized to the transmission of the sapphire substrate by dividing the FFT spectra of the films and substrate by the transmission spectrum of the substrate alone (Supporting Information Figure S6). The averaged THz transmission values in the 0.2- 2THz domain were used to plot the normalized THz transmission vs. temperature or applied current hysteresis curves, similar to the ones obtained in the 4-point probe DC electrical measurements. Finally, the amplitude modulation depth (MD), which describes the modulation of the electric field amplitude, of the devices was computed from MD =[($A_{max}$-$A_{min}$)/$A_{max}$], where $A_{max}$ and $A_{min}$ correspond to the maximal and minimal detected transmitted THz radiation amplitude[39].

Infrared thermal images of the devices during the thermal and electrical activation of the films were taken simultaneously using a Thermal LWIR Optris PI 640i camera with a f 44 mm focal microscope optics.

3. **Temperature-induced transition**

The W percentage of the W:$VO_2$ fabricated films, as analysed by Rutherford backscattering spectroscopy, was revealed to be 0.41 at% for D1, 1.16 at% for D2 and 1.86 at% for D3 samples, respectively. RBS spectra and the corresponding simulations that allow to derive the concentrations of W are given in the Supporting Information Figure S2.

*3.1 Structural and electrical properties of the fabricated films*

All films are epitaxial with the (010) planes of $VO_2$ being parallel to (001) planes of sapphire (see experimental section). The corresponding XRD θ-2θ scans are given in the Supporting Information Figure S3. **Figure 1**a shows the evolution of the 020 XRD peak of the M1 phase upon increasing W doping levels. With increasing W content, the peak shifts towards lower angles indicating a swelling of the unit-cell upon incorporation of W in the $VO_2$ lattice. From Bragg's law, the position of the 020 peak of the M1 can be converted into the corresponding lattice parameter, b = sin(θ)/ λ, and the strain associated to the peak shift hence corresponds to e = (b – $b_0$)/$b_0$, where $b_0$ is the b lattice parameter of the undoped film. It turns out that the strain is linearly dependent on the doping level with a slope of 0.067; i.e., 1 at% tungsten gives rise to a 0.067% increase of the b lattice parameter of $VO_2$. The film with highest W content exhibits a strain of 0.12%. Taking into account the fact that upon the M1-R transition the lattice spacing in this direction increases by ~0.6% [35], W incorporation deforms the unit-cell towards a rutile-like structure which is one of the driving mechanisms of the $T_{IMT}$ reduction.

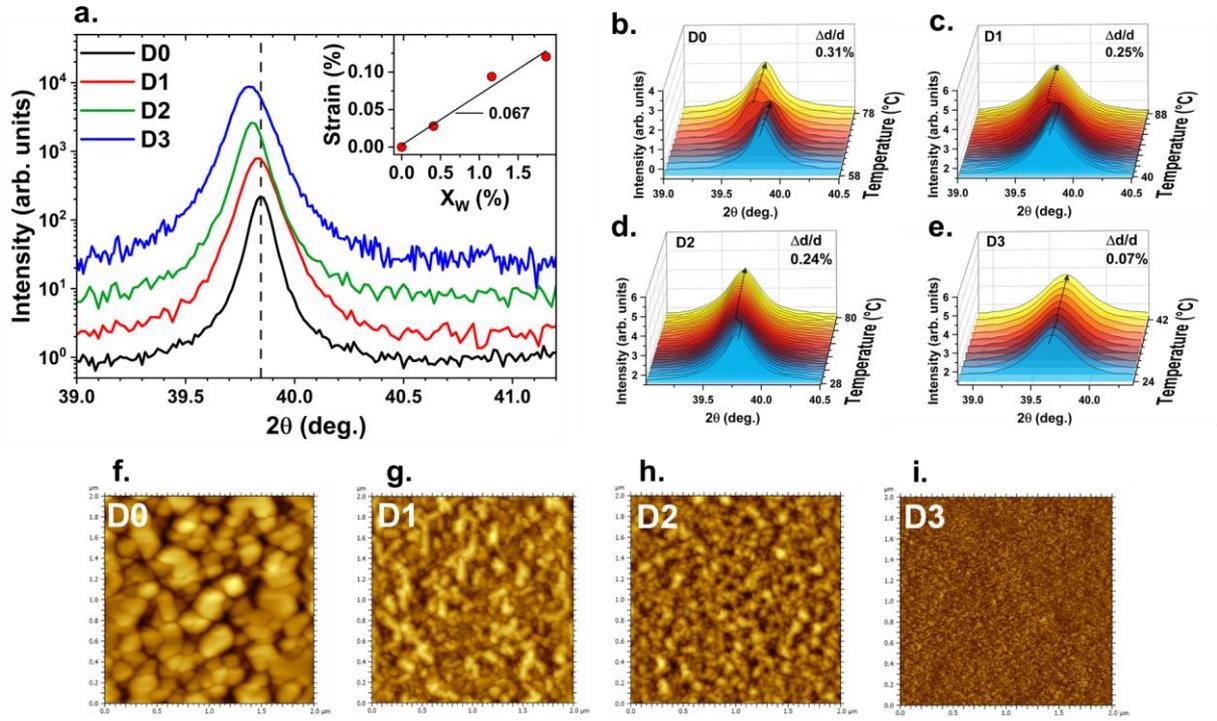

**Figure 1.** a. a) θ-2θ scans of the films with different at.% W doping, of 0% (D0), 0.41% (D1), 1.16% (D2), 1.86% (D3), zoomed on the (020) peak of $VO_2$; the inset graph is tracing the $e_{zz}$ strain for every doping level. b– e. Temperature-dependent evolution of the M1 phase towards the R phase for each of the samples D0-D3. f.-i. the corresponding 2×2 μm² -area AFM scans of the D0-D3 layers.

The temperature-dependent XRD measurement in the vicinity of the 020 Bragg peak of $VO_2$ films with different doping levels are shown in **Figure 1** b-e, where it's evident that, as the doping amount increases, the angular separation between the peaks of the M1 and R phases diminishes. The corresponding relative lattice spacing difference between the two phases $\Delta d/d = (d_R − d_{M1})/d_{M1}$ is indicated in the figure and evolves from 0.31% to 0.07%. Concomitantly, the intensity ratio between the peaks of the R and the M1 phase is decreasing from 1.35 for the D0 sample to 1.14 for D3. These two observations suggest that (i) the unit cell of the two phases are converging to similar dimensions and, (ii) the structure factor are getting similar as well. As outlined in [30]. XRD alone is not sufficient to draw firm conclusions regarding a possible structure change of the VO$_2$ phases. However, in their work, as well as in [27], using pair distribution function analysis and/or X-ray absorption spectroscopy, the authors demonstrated that these observations could be linked to a distortion of the M1 phase towards the R phase.

From the XRD data shown in **Figure 1 b-e** we can derive the volume fraction of the M1 and R phase using the methodology detailed in section 2.3. The corresponding data simulation are shown in the Supporting Information **Figure S4**. The volume fraction evolution of the M1 phase with temperature is displayed in **Figure 2a.**, where a SPT hysteresis is clearly seen. The corresponding parameters of the hysteresis are given in **Table 1**. For increasing doping levels, the average SPT transition temperature decreases from 66.9 to 33.8°C. The hysteresis width varies from 1.9 to 3.4°C, but with large uncertainties (up to 2.1°C) so that no firm conclusion can be drawn for this parameter. The transition width increases from 4.1 to 17°C, which will be further discussed below with regards the grain size of the films. It is worth noting that for D3, the volume fraction of M1 is not 1 at room temperature, indicating the R phase is already present. This is due to the fact that the transition temperature is extremely close to room temperature.

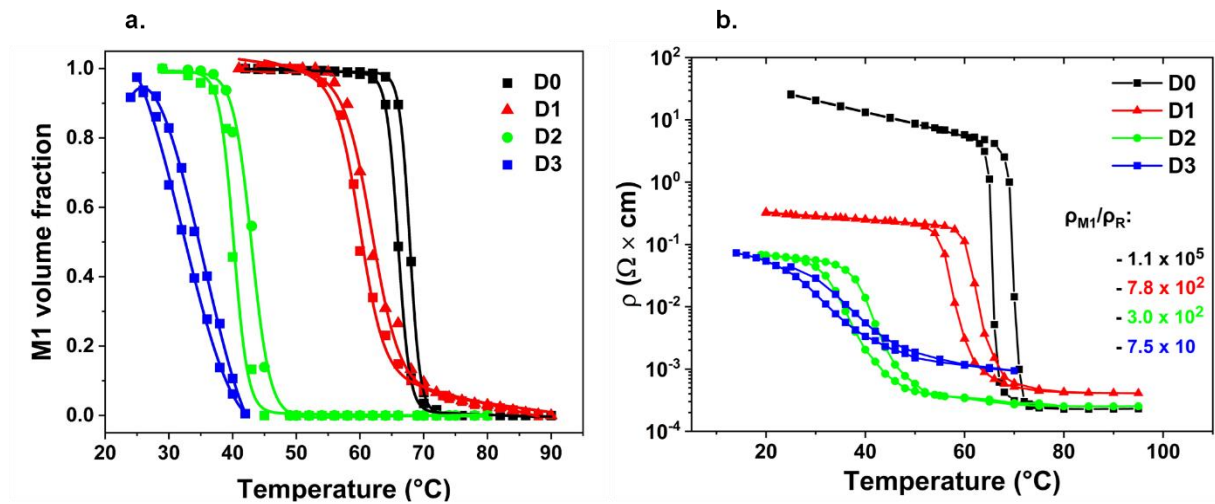

**Figure 2**. Temperature-dependent of the a. M1 phase volume fraction during the structural phase transitions and b. electrical resistivity evolutions of the undoped (D0), 0.41 at.% W (D1), 1.16 at.% W (D2) and 1.86 at.% W (D3) doped layers.

| Doping (at.% W) | $T_{avg}$ (°C) | ΔH (°C) | ΔT (°C) |
|---|---|---|---|
| 0 (D0) | 66.9 ± 0.1 | 1.9 ± 0.1 | 4.1 ± 0.1 |
| 0.41 (D1) | 60.9 ± 0.3 | 1.9 ± 0.7 | 7 ± 0.7 |
| 1.16 (D2) | 41.6 ± 0.3 | 2.8 ± 0.5 | 6.1 ± 0.5 |
| 1.86 (D3) | 33.8 ± 1 | 3.4 ± 2.1 | 17 ± 2.1 |

**Table 1.** Parameters of the SPT hysteresis curves of the films presenting different W-doping percentages.

The Raman spectra of all D0-D3 films, recorded at room temperature, are shown in **Figure S5** of the Supporting Information. All spectra are characteristic of the M1 phase of $VO_2$. Let us first consider the evolution of the D0 sample with increasing temperature (**Figure S5 b**): the Raman peaks intensities start to decrease around 68°C and completely vanish at 70°C, where the signal solely consists of a background signal characteristic of a metallic material, i.e. the R phase. The most doped sample (D3) shows a similar behaviour but with a transition temperature shifted towards a lower value: the peak intensity starts to decrease between 50 and 55°C and the signal totally vanishes at 60°C. It can be noticed that this transition temperature differs from the one observed by XRD. Although that type of discrepancy is not uncommon when comparing thermally-resolved Raman scattering and XRD measurement, the exact origin is not entirely clear. A likely reason might lie in the nature of the probe: whereas XRD is sensitive to long range crystalline order, which requires large regions of the crystal to be transformed to yield a significant signal, Raman spectroscopy is more sensitive to short range order and can hence detect subtle changes in chemical bonding, even if those are randomly distributed throughout the material.

The temperature-dependent electrical resistivity evolution of the films is presented in **Figure 2b**. It can be clearly seen that, as the doping amount increases, so does the shift of the IMT temperature towards lower values, along with the widening the transition (increasing ΔT), as well as the decrease in the resistivity ratio between the insulating and conductive phases. This evolution is in agreement with previously published results regarding the effect of W doping in $VO_2$[27,40]. The parameters quantifying the electrical transition are summarized in **Table 2**. Firstly, an important result is the IMT and the SPT temperatures are identical within experimental uncertainty. Secondly, it can also be observed that, on average, the hysteresis width and the transition width are larger for the IMT. The most likely reason for this discrepancy is the difference in sensitivities between XRD and resistivity measurements. For XRD, when the volume fraction of one of the two phases reaches a very low level, the corresponding intensity also reaches very low values, preventing any quantification, in particular when the peaks of the two phases highly overlap. On the contrary, for the electrical measurement a resistance value is measured whatever the volume fraction of the phases. Thirdly, similarly to what is observed for the SPT, whereas the hysteresis width remains constant within experimental uncertainty, the width of the transition increases with increasing doping level. This is further discussed below.

Lastly, the resistivity ratio, $\rho_{M1}/\rho_R$, follows a monotonic decrease with increasing doping levels, dropping from five orders of magnitude to only one order of magnitude for the highest level of doping. As mentioned before and in previous reports,[27,29,30] this decrease could be explained by the electron excess upon W atom substitution in the local M1 structure. The average IMT temperature decreases from 67.7°C for the undoped film to 34.7°C for the most doped one. This evolution correspond to a linear behaviour with a slope of 18.9 ± 2.9°C/ at% W doping level, which is consistent with previously published results[27] and is attributed to the lower energy barrier between the M1 and R phases for the doped films.

In order to get more insight regarding the evolution of the transition temperature and the evolution of the transition width, we conducted atomic force microscopy (AFM) observation of the surface of the films, **Figure 1 f-i.** It is readily observed that with increasing W content, the average lateral grain size in the films decreases from 207 ± 76 nm (D0) down to 99 ± 29 nm (D2) while for the highest doping level (D3) the grain sizes of the film were too small to be accurately determined. The correlation between decreasing grain sizes and the increase of dopants concentration has been described before and was associated to the increase of density-related nucleation effects during film growth[20]. The increase of the grain boundary density associated with the decrease of the grain size most likely explains the increase of the resistivity level in the conducting phase (similarly to what is observed in metallic films [41,42]) which is another factor explaining the reduction of the resistivity ratio.

Another consequence of the grain size reduction is the increase of the transition width both in the SPT and IMT transitions. This effect results from the combination of the increase in strain level combined with the decrease in grain size. Indeed, the multiplication of the number of grains with different sizes translates to a wider variety of grains with different level of strain, hence a larger distribution of $T_{IMT}$ and $T_{MIT}$ values. Finally, as already noted above, both for the IMT and SPT, the hysteresis width remains roughly constant within the experimental uncertainty. According to Narayan and Bhosle's model[43], in films of high structural quality with large grains separated by low angles, a sharp transition with large amplitude and small hysteresis width should be seen, which is indeed our case.

| **Doping (at.% W)** | $T_{avg}$ (°C) | $\Delta H$ (°C) | $\Delta T$ (°C) | $\rho_{M1}/\rho_R$ ratio |
|---|---|---|---|---|
| 0 (D0) | 67.7 ± 0.1 | 4.2 ± 0.2 | 5.3 ± 0.2 | $1.1 \times 10^5$ |
| 0.41 (D1) | 60.6 ± 0.2 | 4.7 ± 0.4 | 8.9 ± 0.4 | $7.8 \times 10^2$ |
| 1.16 (D2) | 40.0 ± 0.3 | 4.9 ± 0.5 | 12.3 ± 0.5 | $2.7 \times 10^2$ |
| 1.86 (D3) | 34.7 ± 0.3 | 4.6 ± 0.6 | 16.2 ± 0.6 | $7.5 \times 10$ |

**Table 2.** Parameters of the electrical hysteresis curves of the films presenting different W doping percentages.

*3.2 THz modulation*

The evolution of temperature-dependent normalized THz amplitude transmissions of the D0 and D3 samples as a function of frequency are shown in **Figure 3**a and **3**b while the hysteresis of the normalized THz amplitude transmissions (mean values between 0.2 and 2 THz) during a heating- cooling cycle is represented in **Figure 3**c for all samples. In the context of THz applications, the abrupt variation of the transmission around the transition temperature, observed for the D0 sample, is typical of a switching behaviour, as is also evidenced in Figure 3 a where the transmission abruptly drops over the whole spectral range investigated (see also additional Figures S6 a in the Supporting Information). On the contrary, the smooth transition observed for the D3 sample, also visible in **Figure 3**c, is typical of a THz continuous amplitude modulator. This effect has already been observed in previous studies [44,45].

As in the case of the electrical and structural transition, the transition temperature $T_{avg}$ of the THz transmission decreases for each subsequent level of doping (**Figure 3**c) but, interestingly, the magnitude of the transition is not as drastically affected for the THz transmission as compared to the changes in the electrical resistivity presented in **Figure 2**b. Indeed, the THz modulation depth for the samples D0 to D3 only decreases from 96.3% (D0 sample) down to 82.8% (D3 sample), as compared to the ~2000 times drop in electrical resistivity ratio for the same films.

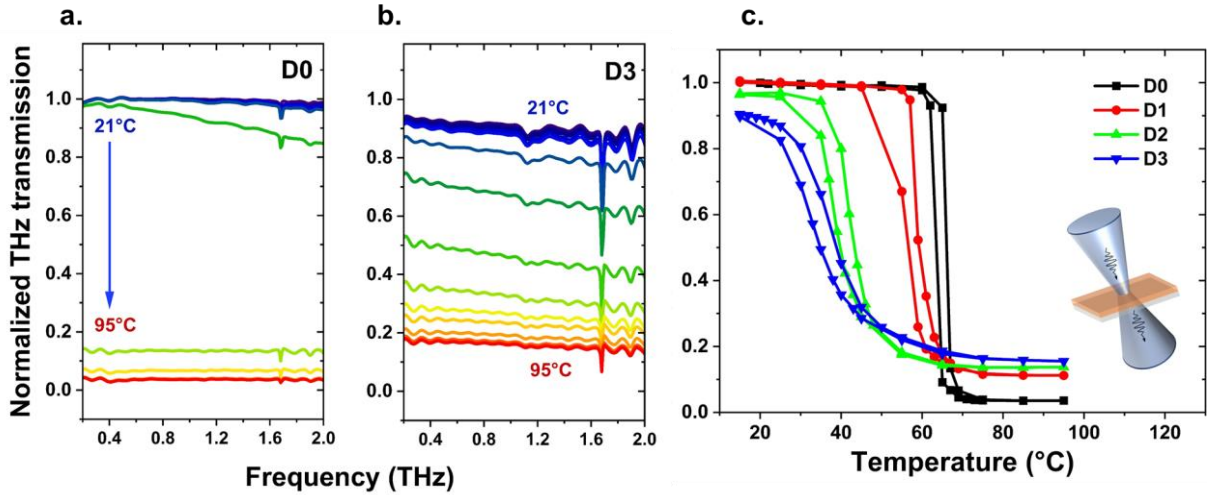

**Figure 3.** Temperature- and frequency- dependent normalized THz transmissions of samples D0 (a.) and D3 (b.) between 0.2 and 2 THz; c. temperature evolution of normalized THz amplitude transmission (mean values between 0.2- 2THz) for samples D0-D3 during a heating-cooling cycle. The inset is sketching the THz measurement configuration in which the substrate and the layers are mounted on a heating Peltier element with a circular hole in the middle, to control the thermal activation and let the THz pulses through.

The modulation depth in the THz domain has been previously noted to be heavily influenced by the metallic phase and the free carrier density of the films [46]. Since W doping increases the electron density in the conduction band, the THz transmission at room temperature is progressively reduced, as observed in **Figure 3**b. However, even though the electrical resistivity level is indeed lowered in the insulating phase of the doped films, it appears that the free electron density is not high enough to drastically affect the transmitted THz amplitude in the investigated frequency domain. Differences between the amplitude evolution of the hysteresis curves within each set of electrical and THz measurements also arise from structural defects (like those at the film/ substrate interface [35]) and grain boundaries, which influence how the phase transition spreads through the material. Whereas structural defects and grain boundaries have a high impact on the electrical measurements by limiting carrier mobility or acting as charge carrier traps, these effects are negligible in the THz measurements which are probing the overall response of the films within the THz beam (~2 mm in diameter). This was previously confirmed in several studies [44], which confirmed the link between the normalized THz transmission of the films for a specific temperature - and hence for a specific conductivity of the films, to the layers' thicknesses and the films' conductivities (equation (S1) and Figure S10 in the supporting information). Using this relation, we obtained a good agreement between the theoretical simulation and the experimental results of the normalized THz transmissions for the D0 and D3 films at different temperatures, which is obvious when comparing the curves in Figure S10, especially at low terahertz frequencies. Thus, the modulation depth of the films (either un-doped or W-doped $VO_2$ films) is directly related to their measured electrical conductivities directly extracted from Figure 2b.

A noticeable difference between the THz and electrical measurements of the films is that the absolute values of $T_{avg}$ are different. It has been observed that, since the electrical transition requires the formation of a conductive path between the electrical probe tips [47], the electrical transition is delayed with respect the optical one, which can be detected as soon as the first nuclei of the R phase are formed [47]. However, such a delay is not systematically observed in our case: whereas the onset of the transition (observed from the point of view of THz radiation) for the D0 and D1 samples is slightly faster, the D2 and D3 samples exhibit the opposite behaviour. A possible explanation could be due to the aperture in the Peltier element which is letting the THz radiation through. A direct consequence of the presence of this aperture is that the area of the film above it is not heated at the same rate compared with the regions in contact with the Peltier element, resulting in temperature variations in this region. In the case of the electrical resistivity measurements the whole films areas are in contact with the heating element.

| Doping (at.% W) | $T_{avg}$ (°C) | ΔH (°C) | ΔT (°C) | MD (%) |
|---|---|---|---|---|
| 0 (D0) | 64.8 ± 0.1 | 2.6 ± 0.2 | 3.8 ± 0.2 | 96.4 |
| 0.41 (D1) | 57.6 ± 0.4 | 3.2 ± 0.8 | 7.1 ± 0.8 | 88.8 |
| 1.16 (D2) | 40.8 ± 0.5 | 3.6 ± 0.9 | 9.6 ± 0.9 | 85.7 |
| 1.86 (D3) | 35.4 ± 0.6 | 4.3 ± 1.2 | 15.4 ± 1.2 | 82.8 |

**Table 3.** Hysteresis parameters of the normalized THz amplitude transmission curves in **Figure 3**c.

For the development of THz modulators, the most interesting feature emerging from these characterizations is the widening of the transition (ΔT) in the case of the doped films, observed both for the electrical resistivity and the THz transmission measurements. While promising, the use of the direct thermal activation is not ideal for functional devices which usually require a form of electrical control over the active elements. This topic is extensively investigated in the next section.

4. **Electrical control of the metal-insulator transition on large-area VO$_2$-based devices**

*4.1. Electrical characteristics of the devices integrated VO$_2$-based films*

For the electrical activation of the different VO$_2$-based layers, we fabricated on top of the D0-D3 films two parallel metallic electrodes made of a Ti/ Au bilayer (30/ 600 nm in thickness) by employing photolithography and metal electron-beam evaporation. The electrodes cover the whole width of the $10 \times 10$ mm$^2$ substrates and leave an exposed area of $10 \times 3.8$ mm$^2$ of the VO$_2$-based films (**Figure 4**a), i.e. wide enough to let the THz beam pass through the device. It is worth noticing that this interelectrode distance is considerably larger than the micrometre-scale distances usually reported in the literature for the electrical activation of MIT of VO$_2$ films [48,49]. Conducting wires were further soldered on the electrodes for convenient integration in the electrical activation and measurement circuit. The electrical activation of the films was performed using a voltage/ current source (TTI with maximum voltage/ current of 120V/ 1.5 A). Alternatively, a Keithley 2612B source-meter was employed for recording the current-voltage (I-V) characteristics of the devices.

For these experiments, the current compliance was set at a value larger than the supposed current threshold for the IMT. The voltage value from the source was then gradually raised until an abrupt decrease was noted, indicating the initiation of the IMT, as noted on the I-V curves represented on **Figure 4** b for the samples D0 and D3. Afterwards, the current compliance was further gradually increased, usually with increments of 50 mA while the IR thermal imaging was used to assess the metal and insulating phases. In the thermal images of **Figure 4** the top row (c-i) correspond to the D0 sample, while the second row (j-p) correspond to the D3 sample. In both cases, the leftmost image (c and j) correspond to the case where no current is applied, while the following images correspond to different current values: d and k correspond to the current value at the onset of the transition; i and p correspond to the maximum applied current.

The electrical activation of the D0 device, was achieved with the sample being kept at temperatures between 57°C and 64°C, i.e. slightly below the IMT, in order to reduce the threshold voltage/current needed for electrically-induced IMT, as voltage and current threshold values are decreasing for higher bias temperatures. When the resistance of the D0 device dropped due to the onset of the electrically induced IMT, the thermal camera registered the appearance of a channel at a seemingly lower temperature (**Figure 4** d-h). This apparent cooling is in fact due to formation of the metallic $VO_2$ channel which has a lower emissivity than the insulating $VO_2$. For these large area devices, the electrical activation is most likely purely thermal, through Joule heating. Under the effect of temperature, metallic nuclei appear at the onset of the transition, and coexist with the insulating phase until percolation is reached, which occurs through the formation of a filament that creates a conductive path between the electrodes. As the applied current is further increased, the metallic channel grows laterally until the whole sample convert to the metallic state for temperatures higher than the IMT transition temperature[50,51]. It must be emphasized that, in the present case, the narrow state of the filament is not observed because, for increasing voltage, the current abruptly jumps from low to high values, leading the filament to rapidly reach a broad size (**Figure 4** d). It is important to mention that upon several transition cycles, the filament randomly appears at different locations (Supporting Information Figure S7), confirming the stochastic nature of the nucleation/percolation process.

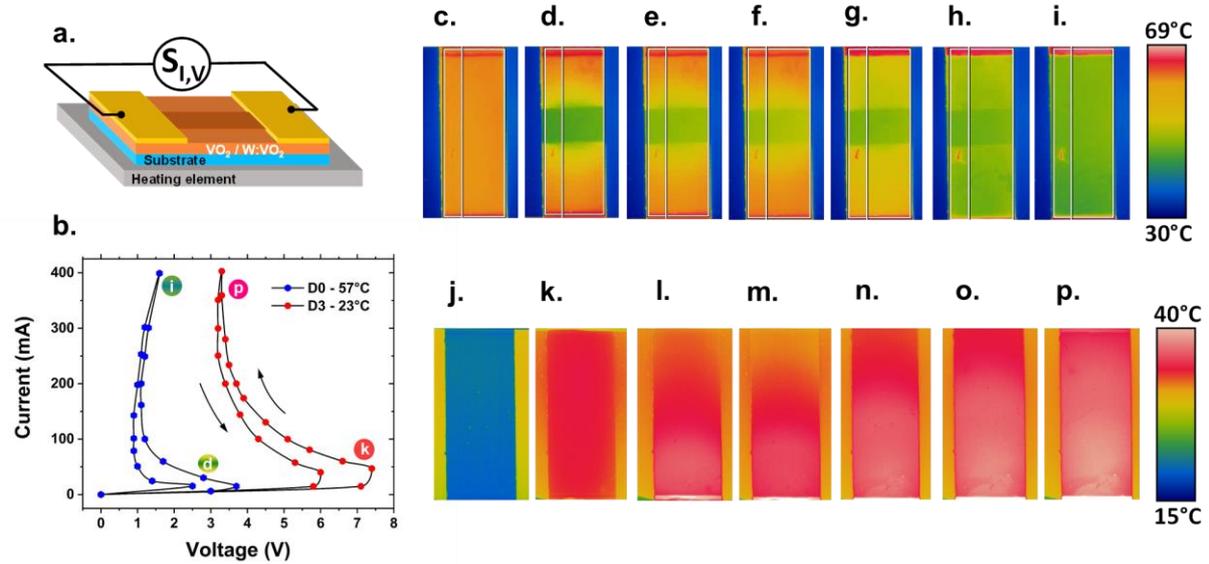

**Figure 4**. a. Schematics of the devices fabricated for the electrical control of the IMT in $VO_2$-based layers integrated with a current-voltage electrical source, b. typical I-V characteristics of the D0 and D3-based devices showing a negative differential resistance S-shape and IR thermal images at different stages of metallic channel expansion of the undoped D0 film - based device at a bias temperature of 64°C (c -i) and of the most doped D3 films - based device at a temperature of 23°C during current increasing between 0 and 400 mA (j- p).

While a similar behaviour can still be observed for D1, overall, for increasing doping level, the onset of the metallic channel becomes less visible, until it completely disappears for the device integrating the most-doped sample, D3, **Figure 4** j-p. For this specific case, the D3 device was kept at a temperature of 23°C in order to ensure that all the domains are in the M1 phase. At a current of 50 mA the film has transitioned to the metallic state although no channel can be detected. In this case, because of the structural similarity of the two phases and, hence, the lower energy barrier between the two states, the metallic-insulating nanodomains mixture is much closer to the percolation threshold than in the case of the D0 device; i.e. the amount of Joule heating required to transform an M1 grain into the R phase is much lower than for the undoped sample. This leads the electrical induced transition to develop in a much more diffuse way, over the entire surface between the two electrodes, in a rather non-localized way. This effect is also likely enhanced by the fact the grains in the D3 samples are smaller than in the undoped case, hence facilitating the switching and subsequent interactions between domains because of the increased grain boundary area. Further refined characterization at micro- and nano-scales, for D0 and D3 devices with different areas and distances between the electrodes will be necessary in order to fully assess this dissimilar electrical IMT activation mechanism in W-doped samples. The next section is the first step in this direction, where we use *in situ* spatially resolved XRD to characterize the crystalline structure the conductive and insulating filaments, during voltage-controlled activation of the IMT.

Finally it is worth mentioning that a phase change purely triggered by the effect of electric field, without Joule heating, has sometimes been reported but for much smaller devices with micron-scale distance between the planar electrodes [52]

*4.2 Spatial distribution of the electrically induced structural phase transition*

The spatial distribution of the SPT during electrical activation of IMT in the devices was investigated using the set-up displayed in Figure 5a. The linear X-ray beam is positioned perpendicular to the electrodes. The detector position is fixed either at the 2θ Bragg angle of the (200) planes of the R phase or the (020) planes of the M1 phase, while the incidence angle is set at the corresponding θ angle. The angular opening Δ2θ of the detector was set to 0.01°, which allows to clearly separate the contribution of the M1 and R phases. With the incidence angle of the X-ray beam corresponding to the 020 reflection of $VO_2$, and considering the beam cross-section size (h = 200 μm), the beam footprint on the sample is S ~ 600 μm. Within these conditions, using the x translation, the X-ray beam was scanned across the sample with steps of 250 µm in the direction parallel to the electrodes.

**Figure 5**b and **Figure 5**c represent the intensity of the 020 peak of the M1 phase at different x positions with respect to the centre of the device during the current-controlled IMT for the undoped (D0) and the most doped films (D3). For D0, when no voltage is applied (red curve) the intensity of the scan is roughly constant, indicating that the entire layer is in the M1 phase, thus insulating. Actually, a slight increase in intensity (in the region corresponding to positive scanning distances) can be observed. This effect is most likely due the fact that we are scanning a region of the film where the thickness is not perfectly constant (i.e. close to an edge). This evolution is different from sample to sample and does not reflect the properties of the SPT. Moreover, the magnitude of this effect is negligible as compared to the effect induced by the SPT discussed below.

For the first applied voltage (corresponding to a current of 27 mA, dark orange curve in **Figure 5**b) a dip is observed in the scan, evidencing the disappearance of the M1 to the benefit of the R phase. This shows that the conducting filament correspond to the R phase. The scan on the peak of the R phase is given in the Supporting Information and exhibits the opposite behaviour which confirms this statement (Supporting Information **Figure S8**). The boundary between the low and the high intensity region of the curve hence corresponds to the boundary between the M1 and R phase, i.e. to the edge of the electrically induced metallic channel. In the present case, it turned out that the channel emerged at one extremity of the sample. Upon progressively increasing the current up to 350 mA, the boundary clearly moves towards smaller position values along the scanning line, which demonstrates that the colour change observed in the IR measurement is indeed related to the SPT / IMT. For the highest current, almost all the film is converted to the R phase.

Also provided in Figure 5 are θ-2θ scans recorded for low, and high x values, for all the different applied voltages. These scans demonstrate that at x = 3mm the M1 phase rapidly disappears at the expense of the R phase, whereas at x = -3 mm, the M1 phase persist up to high currents.

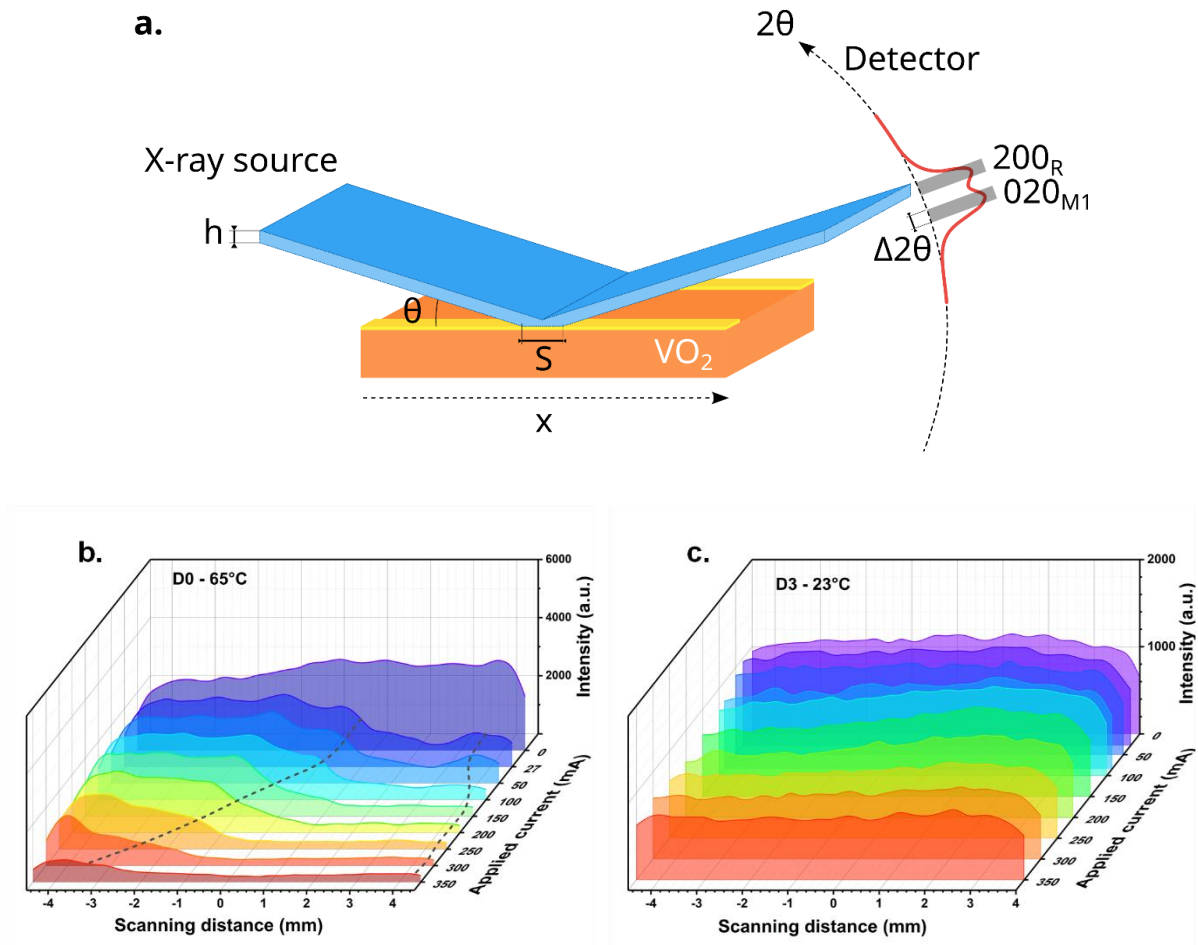

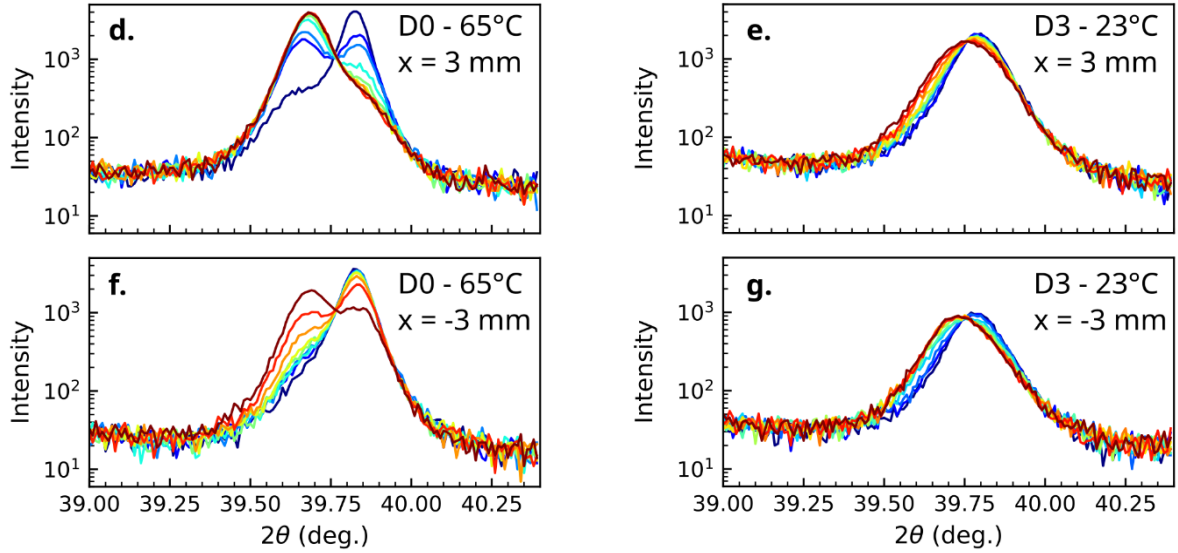

**Figure 5.** (a.) XRD setup used to spatially map the VO$_2$ phases during the electrically activated transition. (b.) XRD spatial mapping of the M1 phase on a direction parallel to the device's electrodes, showing the evolution of the in the D0 film kept at 65°C (c.) and in the D3 film at 23°C, as a function of the applied current between the electrodes. (d.- f.) θ-2θ scans from the D0 film recorded at x = ± 3 mm. These scans demonstrate that at x = 3mm the M1 phase rapidly disappears at the expense of the R phase, whereas at x = -3 mm, the M1 phase persist up to high currents. (e,g) same as before for the D3 film. In this case, there is no clear difference between the two positions.

In striking contrast, a similar measurement performed on the most W-doped VO$_2$ sample shows no evidence of a M1/R boundary, but rather a uniform decrease of the intensity with increasing applied current along the whole scanned line. This evidences that, in agreement with the previous IR thermal investigations, the whole film is progressively, but homogeneously transformed, without the formation of a conductive filament. This is also confirmed by the θ-2θ scans shown is **Figure 5** which demonstrates that the behaviour of the transition is identical at both extremities of the film. This is in line with the result presented in the temperature-dependent XRD measurements in **Figure 1**e. and the temperature-dependent electrical resistivity measurements **Figure 2**b. of section 3.1, where the drop in the $T_{avg}$ value was attributed to a lowering of the energy barrier between the two phases

*4.3 Discussion*

**Figure 3**c shows that, in the 0.2 – 2 THz frequency range, the behaviour of $VO_2$ can be tuned from an abrupt switching behaviour to a smooth THz amplitude modulation by increasing the W content. As shown above, it is possible to electrically activate the transition of $VO_2$ layers on areas as large as $10 \times 3.8$ mm$^2$. However, in such a case, and for low level of doping, the transition does not affect the whole film uniformly, but instead proceeds by the nucleation and subsequent growth of a conducting channel. This feature hinders the use of $VO_2$ as THz switches and or/ modulators, since the overall transmission of the film is then determined by the lateral growth of the conductive channel along with its initial position and dimension against the incident THz beam.

Notably, the W doping, besides lowering the IMT close to room temperature, which is in general the primary reason for which it is used, also increases the structural similarity between the M1 and the R phase. This structural distortion, which is at the origin of the drop in the IMT temperature, nonetheless has detrimental consequences on the IMT itself, in particular by decreasing the resistivity ratio by around 3 orders of magnitude. However, the THz transmission is not as much affected as the resistivity since the temperature-induced amplitude modulation only drops from 96.3 to 82.8%.

Thus, a remarkable feature of $VO_2$ doping with W at 1.86 at.% is that the transition occurs homogeneously across the sample, without the formation of a metallic filament. This characteristic allows to envision the development of W:$VO_2$-based, electrically activated amplitude switches and modulators, without the need of any heating bias.

The demonstration of this effect is shown in **Figure 6**. In these experiments the THz transmission of the D0 (kept at 57°C) and D3 films (close to room temperature, 23°C) was monitored as a function of the applied current across the device during ten hysteresis cycles (current increasing and decreasing between 0 and 400 mA with steps between 25 and 50 mA) which are superposed on the graphs in **Figure 6** for both D0 and D3 devices.

As indicated on **Figure 6**a and **Figure 6**b, the THz transmission is exhibiting a neat and continuous decrease with increasing the applied current, corresponding to a MD of 92.2% and 85.4%, for D0 and D3, respectively. For D0, the evolution of the transmission is not perfectly reproducible across different cycles. Moreover, the transmission exhibits a steep drop in transmission and a marked hysteresis. In striking contrast, D3 is showing outstanding repeatability across the ten measurements performed back-to-back, with almost no noticeable hysteresis between the increasing and decreasing branches.

The unpredictable behaviour of D0 can be explained by the aforementioned random position of the metallic channel relative to the THz beam position in the undoped D0 layer and the lateral dimension of this filament compared to the THz beam (see inset sketch in **Figure 6**a). As previously shown, for the D3 device the transition take place evenly across the sample and within the THz beam, with no metallic filament formation (sketch in **Figure 6**b), which explains the excellent reproducibility of the measurements.

Besides the MD, an interesting metric to consider is the spread of the modulation in the THz transmission– current quadrant (grey areas in **Figure 6**a and **Figure 6**b. In order to quantify this behaviour we introduce the modulation range (MR) metrics, defined as:

$$MR = \frac{A_{95\%} - A_{5\%}}{I_{95\%} - I_{5\%}} \qquad (3)$$

where $A_{95\%}$ and $A_{5\%}$ are the THz transmissions amplitudes at 95% and 5% with respect to the maximum amplitude and the $I_{95\%}$ and $I_{5\%}$ are the corresponding current values at which these values are occurring. As indicated on **Figure 6**a and **Figure 6**b, the MR value of the D0 device is roughly two times higher than that of D3, indicating a much sharper variation of the THz transmitted amplitudes with the applied current. In contrast, the wider MR of D3 points to a wider range of transmission operating points for the same applied current range.

In addition, the D3 device was submitted to more than 38.000 periodic cycles of square current pulses (300 mA amplitude, pulse length and period of 500 ms/1000 ms) in order to record the long-term stability of the MD during its electrical activation. The constant evolution of the MD with the number of cycles shown in **Figure 6 c.** demonstrates the very good reliability of the device performing a large number of insulator-to metallic activation cycles.

Thus, the performances of the D3 device integrating the W:VO$_2$ layer with the highest concentration is surpassing the ones of the bare VO$_2$ based device in terms of reliability, reproducibility and modulation range, while operating over a broad THz range at room-temperature, with electrical control over the largest active area reported so far for THz modulators (10 × 3.8 mm$^2$). In addition, the activation time of the D3 device was evaluated by applying a square-shape current pulse over the device and recording its electrical response (Figure S9). The response time to the applied current pulse was as low as 24 ms while the insulator-to-metal transition occurred within 339 ms, which is remarkable for such large-area devices and the employed electrical activation scheme.

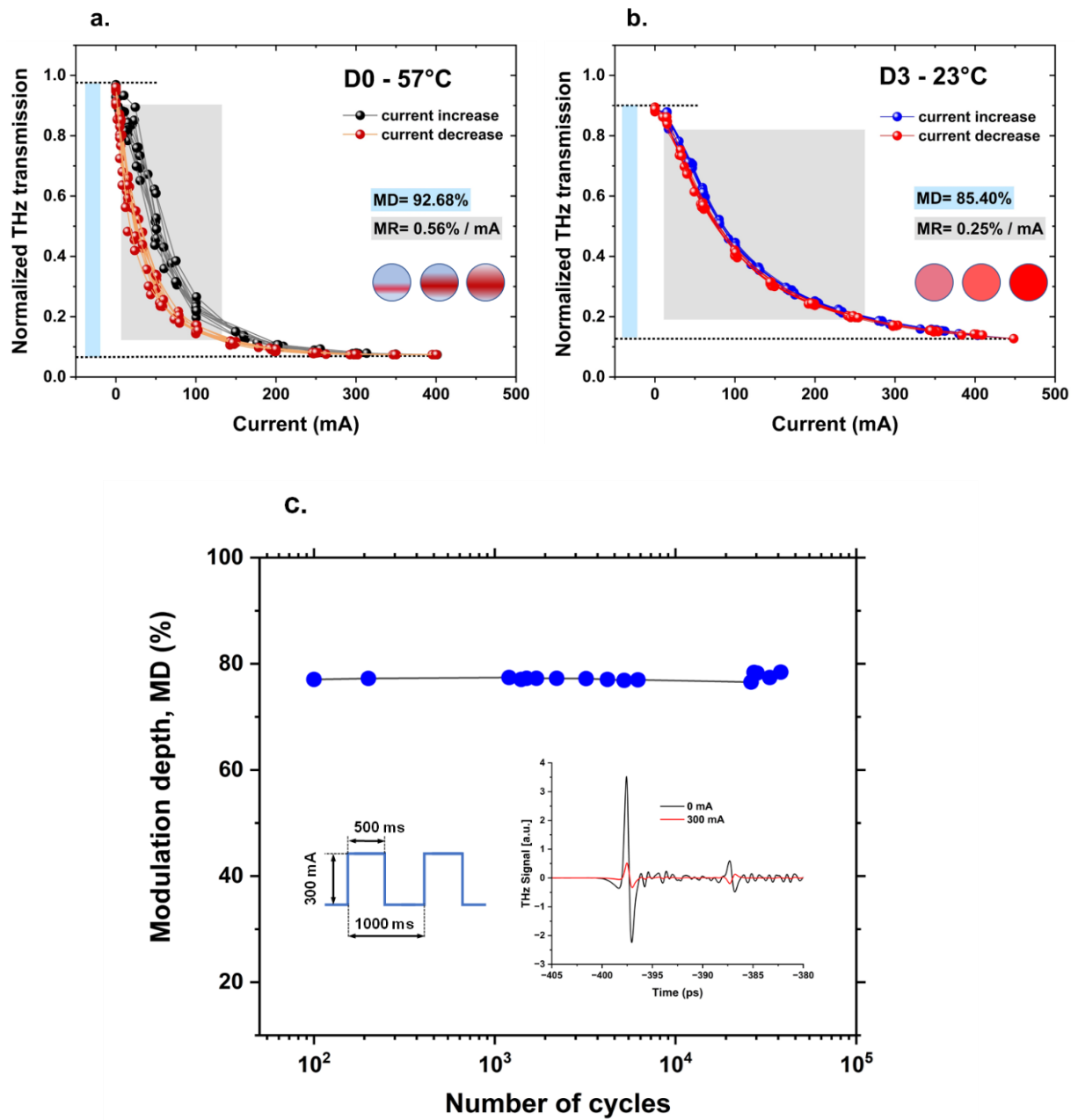

**Figure 6**. Superposition of ten consecutive current-dependent THz transmission curves (0.2 – 2 THz) for the a. D0 device kept at 57°C and b. the D3 device at 23°C, indicating the modulation depth (MD) and the modulation range (MR) metrics for each device. The inset sketches are illustrating the different electrical activation behaviour of the two materials within the THz beam. c. Investigation of long-term stability of the D3 device showing constant THz modulation depth after more than 38.000 cycles when periodically switching the device using a current square waveform with 300 mA amplitude, a pulse duration of 500 ms and a period of 1000 ms;

the inset shows the transmitted temporal THz trace of the device in the insulating state (0mA applied current) and in the metallic state (300 mA applied current).

Compared to the D0 device showing an abrupt variation of the THz transmission around the transition temperature and a limited current modulation spread typical to a switching-like behavior, the smoother transition observed for the D3 sample for both thermal and electrical activations, makes its integration highly interesting for applications where a precise and continuous (or discrete) thermally or electrically-triggered amplitude modification is required. The higher modulation range (**Figure 6 b**), the good MD and large area electrically activation along with the stability and reliability of the device are significant metrics for future integration in THz reconfigurable devices such as continuous / discrete attenuators, precise temperature sensing devices, polarizers with chosen polarization configurations or filter-like reflection/ transmission metasurfaces with controlled absorptions and bandwidths, THz wavefront engineering using coding metamaterials etc. While the room temperature operation is a strong feature of the D3-like devices, their lower activation temperature may hinder their large-scale integration for industrial applications but this can be circumvented by integrating the device with thermal bath devices (miniature Peltier or cryogenic) allowing to keep it in a thermal bath much lower than its transition temperature.

We further compared in **Table 4** the performances of the $VO_2$ –based devices investigated here with a selection of relevant THz amplitude modulators and devices reported in the available literature. Although the graphene-based modulator integrated with a quantum cascade laser (QCL) developed in [53] is showing impressing performances, the electrical modulation scheme is rather complex and the device can be employed only at the single THz emission frequency of the QCL source. The use of semiconductor-based THz modulators based on Si or hybrid combinations with transition metal dichalcogenides such as $WS_2$ [54] are based on complex fabrication steps which are critical to their performances and make difficult their industrial upscaling and are usually employing an external high energy density laser source.

Comparing with other similar reported $VO_2$-based modulators, our devices can be employed using thermal and/ or electrical type stimuli. Moreover, the electrical activation scheme implemented here allows the operation of the devices over a large active area, while preserving their broadband frequency capabilities. It can also be noticed that the D0-type device (based on bare $VO_2$ films) perform much better in term of MD, even when compared to specific designs meant to improve the modulation depths [55].

Most importantly, the W:$VO_2$-based D3 device, even if presenting a slightly lower modulation depth, can operate over large interelectrode areas under electrical actuation at ambient temperature and is surpassing the characteristics of D0 device in terms of reproducibility and modulation range. These properties are instrumental for future development and integration of $VO_2$-based layers in THz modulators or meta-devices with state-of-the-art performances.

| Device (Ref.) | Activation trigger | Modulation depth | Frequency (THz) | Modulation speed |
|---|---|---|---|---|
| $WS_2$/Si [54] | Optical | ~99% | 0.25-2 | - |
| Quantum cascade laser integrated graphene modulator [53] | Electrical | 94-100% | 3 THz (fixed single frequency) | < 100 MHz |
| $VO_2$ Metasurface [7] | Electrical | >60% | 0.3- 1 | - |
| Bowtie antenna $VO_2$ metasurface [56] | Electrical | <90% | 0.3- 2.5 | 1 s |
| Hybrid $VO_2$ metasurface [55] | Electrical | 68% | 0.1- 0.8 | 2.1 s |
| Bare $VO_2$ [17] | Thermal | 85% | 0.1- 2 | - |
| Bare $VO_2$ [20] | Thermal | 69% | 4.8- 5.4 | - |
| Bare $VO_2$ [44] | Thermal | 74% | 0.1- 1.5 | - |
| Bare $VO_2$ [57] | Optical | 81.2% | 1 THz | 120 ms |
| **This work** | **Thermal and** | **96.4% / 92.6%** | **0.2- 2** | **-** |
| VO$_2$ | Electrical | | | |
| **This work**<br>**W:VO$_2$** | **Thermal and Electrical** | **82.8% / 85%** | **0.2- 2** | **339 ms*** |

*) see Figure S9 in the Supporting Information

**Table 4.** Comparison of performances reported for a selection of THz modulation devices along with those of the devices presented in this work

5. **Conclusion**

In this work we have grown W-doped VO$_2$ films, with W atomic content ranging between 0 and 1.86 at%. The effect of doping on the structural, electric and THz characteristics have been systematically investigated upon thermal cycling across the corresponding transitions. Overall, with increasing W content, the transition temperature is shifted towards room temperature at a rate of 18.9°C/at% of W doping. XRD characterizations coupled with Raman spectroscopy revealed that this shift in temperature is most likely due to distortions affecting the VO$_2$ M1 and R phases, making them structurally similar to each other. Despite this structural similarity, a clear transition can be observed in terms of structural phases, electrical resistivity and THz transmission under applied temperature. Interestingly, whereas the electrical resistivity ratio between the insulating and conductive phase is significantly altered by W doping, with the ratio dropping by more than 3 orders of magnitude, the THz transmission of the layers is only weakly affected, with a modulation depth dropping from 96.3% to 82.8%.

Electrical activations of the IMT transition on large area undoped and W-doped films were also carried out. For low level of doping, IR thermal observations revealed that the IMT takes place inhomogeneously by the nucleation and expansion of a conductive channel between the electrodes. Spatially resolved operando XRD measurements demonstrated that the metallic channel corresponds to the conductive R phase. These experiments also allowed to demonstrate that, for higher doping levels, the transition occurs homogeneously across the film without the formation of a conductive channel. Such devices using VO$_2$ films doped with 1.86 at% W and large 10 × 3.8 mm$^2$ active area reached a total THz modulation depth of 85.4% at room temperature using solely current injection with intensities ranging between 0 and 400 mA.

This last proof of concept opens important perspectives regarding the fabrication of electrically controlled optical modulators incorporating VO$_2$. Combining multiple type of stimuli can also be an efficient way to take advantage of the large areas over which VO$_2$ can be activated for various applications, such as electrical and optical switches/ modulators in the IR or THz range, thermal regulation, or multi-stimuli reconfigurable optically-written meta-devices.

ASSOCIATED CONTENT

Temperature-dependent resistivity measurement of D2 sample, RBS spectra, θ-2θ scans of the films between 15° and 95°, fit of the XRD temperature-dependent data with a model based on two pseudo-Voigt functions for all D0-D3 samples, RT Raman spectra of all films and temperature-dependent Raman spectra of D3 and D0, THz time-domain and frequency-domain spectra of D0-D3 samples, IR images of D0, XRD scans of the samples along the devices, Voltage response across the D3 device when applying a square-shaped current pulse for IMT response evaluation simulation and experimental comparison of the normalized THz transmissions for D0 and D3 samples at different temperatures.

AUTHOR INFORMATION

**Corresponding Authors**

*Corresponding authors:

**Eduard-Nicolae Sirjita** - XLIM Research Institute, CNRS/ University of Limoges, France, Institut de Recherche sur les Céramiques (IRCER), CNRS UMR 7315, University of Limoges, France; Email: eduard.sirjita@xlim.fr,

**Alexandre Boulle** - Institut de Recherche sur les Céramiques (IRCER), CNRS UMR 7315, University of Limoges, France; Email : alexandre.boulle@cnrs.fr,

**Aurelian Crunteanu** - XLIM Research Institute, CNRS/ University of Limoges, France; Email: aurelian.crunteanu@xlim.fr

**Authors**

**Jean-Christophe Orlianges** - XLIM Research Institute, CNRS/ University of Limoges, France

**Richard Mayet** - Institut de Recherche sur les Céramiques (IRCER), CNRS UMR 7315, University of Limoges, France

**Aurélien Debelle** - IJCLab, Université Paris-Saclay, CNRS/IN2P3, Orsay, France.

**Lionel Thomé** - IJCLab, Université Paris-Saclay, CNRS/IN2P3, Orsay, France.

**Maggy Colas** - Institut de Recherche sur les Céramiques (IRCER), CNRS UMR 7315, University of Limoges, France

**Julie Cornette** - Institut de Recherche sur les Céramiques (IRCER), CNRS UMR 7315, University of Limoges, France

**Notes**

The authors declare no competing financial interest.

ACKNOWLEDGMENT

This work benefited from government support managed by the National Research Agency under the Investments for the future program with the reference ANR-10-LABX-0074-01 Sigma-LIM. The authors acknowledge support from the PETACom FET Open H2020 grant number 829153 and the Femto-VO2 Project (Nouvelle-Aquitaine region).

References

(1) Dhillon, S. S.; Vitiello, M. S.; Linfield, E. H.; Davies, A. G.; Hoffmann, M. C.; Booske, J.; Paoloni, C.; Gensch, M.; Weightman, P.; Williams, G. P.; Castro-Camus, E.; Cumming, D. R. S.; Simoens, F.; Escorcia-Carranza, I.; Grant, J.; Lucyszyn, S.; Kuwata-Gonokami, M.; Konishi, K.; Koch, M.; Schmuttenmaer, C. A.; Cocker, T. L.; Huber, R.; Markelz, A. G.; Taylor, Z. D.; Wallace, V. P.; Axel Zeitler, J.; Sibik, J.; Korter, T. M.; Ellison, B.; Rea, S.; Goldsmith, P.; Cooper, K. B.; Appleby, R.; Pardo, D.; Huggard, P. G.; Krozer, V.; Shams, H.; Fice, M.; Renaud, C.; Seeds, A.; Stöhr, A.; Naftaly, M.; Ridler, N.; Clarke, R.; Cunningham, J. E.; Johnston, M. B. The 2017 Terahertz Science and Technology Roadmap. *J. Phys. D: Appl. Phys.* **2017**, *50* (4), 043001. https://doi.org/10.1088/1361-6463/50/4/043001.

(2) Pang, X.; Ozolins, O.; Jia, S.; Zhang, L.; Schatz, R.; Udalcovs, A.; Bobrovs, V.; Hu, H.; Morioka, T.; Sun, Y.-T.; Chen, J.; Lourdudoss, S.; Oxenlowe, L.; Popov, S.; Yu, X. Bridging the Terahertz Gap: Photonics-Assisted Free-Space Communications From the Submillimeter-Wave to the Mid-Infrared. *J. Lightwave Technol.* **2022**, *40* (10), 3149–3162. https://doi.org/10.1109/JLT.2022.3153139.

(3) Lu, C.; Lu, Q.; Gao, M.; Lin, Y. Dynamic Manipulation of THz Waves Enabled by Phase-Transition VO2 Thin Film. *Nanomaterials* **2021**, *11* (1), 114. https://doi.org/10.3390/nano11010114.

(4) Degl'Innocenti, R.; Lin, H.; Navarro-Cía, M. Recent Progress in Terahertz Metamaterial Modulators. *Nanophotonics* **2022**, *11* (8), 1485–1514. https://doi.org/10.1515/nanoph-2021-0803.

(5) Ma, Z. T.; Geng, Z. X.; Fan, Z. Y.; Liu, J.; Chen, H. D. Modulators for Terahertz Communication: The Current State of the Art. *Research* **2019**, *2019*, 2019/6482975. https://doi.org/10.34133/2019/6482975.

(6) Ren, Z.; Xu, J.; Liu, J.; Li, B.; Zhou, C.; Sheng, Z. Active and Smart Terahertz Electro-Optic Modulator Based on VO$_2$ Structure. *ACS Appl. Mater. Interfaces* **2022**, *14* (23), 26923–26930. https://doi.org/10.1021/acsami.2c04736.

(7) Zhou, G.; Dai, P.; Wu, J.; Jin, B.; Wen, Q.; Zhu, G.; Shen, Z.; Zhang, C.; Kang, L.; Xu, W.; Chen, J.; Wu, P. Broadband and High Modulation-Depth THz Modulator Using Low Bias Controlled VO_2-Integrated Metasurface. *Opt. Express* **2017**, *25* (15), 17322. https://doi.org/10.1364/OE.25.017322.

(8) Parrott, E. P. J.; Han, C.; Yan, F.; Humbert, G.; Bessaudou, A.; Crunteanu, A.; Pickwell-MacPherson, E. Vanadium Dioxide Devices for Terahertz Wave Modulation: A Study of Wire Grid Structures. *Nanotechnology* **2016**, *27* (20), 205206. https://doi.org/10.1088/0957-4484/27/20/205206.

(9) Liu, K.; Lee, S.; Yang, S.; Delaire, O.; Wu, J. Recent Progresses on Physics and Applications of Vanadium Dioxide. *Materials Today* **2018**, *21* (8), 875–896. https://doi.org/10.1016/j.mattod.2018.03.029.

(10) Shi, R.; Shen, N.; Wang, J.; Wang, W.; Amini, A.; Wang, N.; Cheng, C. Recent Advances in Fabrication Strategies, Phase Transition Modulation, and Advanced Applications of Vanadium Dioxide. *Applied Physics Reviews* **2019**, *6* (1), 011312. https://doi.org/10.1063/1.5087864.

(11) Morin, F. J. Oxides Which Show a Metal-to-Insulator Transition at the Neel Temperature. *Phys. Rev. Lett.* **1959**, *3* (1), 34–36. https://doi.org/10.1103/PhysRevLett.3.34.

(12) Beaumont, A.; Leroy, J.; Orlianges, J.-C.; Crunteanu, A. Current-Induced Electrical Self-Oscillations across out-of-Plane Threshold Switches Based on VO$_2$ Layers Integrated in Crossbars Geometry. *Journal of Applied Physics* **2014**, *115* (15), 154502. https://doi.org/10.1063/1.4871543.

(13) Sirjita, E.-N.; Boulle, A.; Orlianges, J.-C.; Mayet, R.; Crunteanu, A. Structural and Electrical Properties of High-Performance Vanadium Dioxide Thin Layers Obtained by Reactive Magnetron Sputtering. *Thin Solid Films* **2022**, 8.

(14) Chang, T.; Cao, X.; Dedon, L. R.; Long, S.; Huang, A.; Shao, Z.; Li, N.; Luo, H.; Jin, P. Optical Design and Stability Study for Ultrahigh-Performance and Long-Lived Vanadium Dioxide-Based Thermochromic Coatings. *Nano Energy* **2018**, *44*, 256–264. https://doi.org/10.1016/j.nanoen.2017.11.061.

(15) Houska, J. Design and Reactive Magnetron Sputtering of Thermochromic Coatings. *Journal of Applied Physics* **2022**, *131* (11), 110901. https://doi.org/10.1063/5.0084792.

(16) Han, C.; Parrott, E. P. J.; Humbert, G.; Crunteanu, A.; Pickwell-MacPherson, E. Broadband Modulation of Terahertz Waves through Electrically Driven Hybrid Bowtie Antenna-$VO_2$ Devices. *Sci Rep* **2017**, *7* (1), 12725. https://doi.org/10.1038/s41598-017-13085-w.

(17) Zhao, Y.; Hwan Lee, J.; Zhu, Y.; Nazari, M.; Chen, C.; Wang, H.; Bernussi, A.; Holtz, M.; Fan, Z. Structural, Electrical, and Terahertz Transmission Properties of $VO_2$ Thin Films Grown on c-, r-, and m-Plane Sapphire Substrates. *Journal of Applied Physics* **2012**, *111* (5), 053533. https://doi.org/10.1063/1.3692391.

(18) Brückner, W.; Moldenhauer, W.; Wich, H.; Wolf, E.; Oppermann, H.; Gerlach, U.; Reichelt, W. The Range of Homogeneity of VO2 and the Influence of the Composition on the Physical Properties. II. The Change of the Physical Properties in the Range of Homogeneity. *Phys. Stat. Sol. (a)* **1975**, *29* (1), 63–70. https://doi.org/10.1002/pssa.2210290107.

(19) Quackenbush, N. F.; Paik, H.; Wahila, M. J.; Sallis, S.; Holtz, M. E.; Huang, X.; Ganose, A.; Morgan, B. J.; Scanlon, D. O.; Gu, Y.; Xue, F.; Chen, L.-Q.; Sterbinsky, G. E.; Schlueter, C.; Lee, T.-L.; Woicik, J. C.; Guo, J.-H.; Brock, J. D.; Muller, D. A.; Arena, D. A.; Schlom, D. G.; Piper, L. F. J. Stability of the M2 Phase of Vanadium Dioxide Induced by Coherent Epitaxial Strain. *Phys. Rev. B* **2016**, *94* (8), 085105. https://doi.org/10.1103/PhysRevB.94.085105.

(20) Pallares, R. M.; Su, X.; Lim, S. H.; Thanh, N. T. K. Optimization of Metal-to-Insulator Phase Transition Properties in Polycrystalline VO2 Films for Terahertz Modulation Applications by Doping. *J. Mater. Chem. C* **2016**, *4* (1), 53–61. https://doi.org/10.1039/C5TC02426A.

(21) Wu, Y.; Fan, L.; Chen, S.; Chen, S.; Chen, F.; Zou, C.; Wu, Z. A Novel Route to Realize Controllable Phases in an Aluminum ($Al^{3+}$)-Doped VO2 System and the Metal–Insulator Transition Modulation. *Materials Letters* **2014**, *127*, 44–47. https://doi.org/10.1016/j.matlet.2014.04.094.

(22) Victor, J.-L.; Gaudon, M.; Salvatori, G.; Toulemonde, O.; Penin, N.; Rougier, A. Doubling of the Phase Transition Temperature of $VO_2$ by Fe Doping. *J. Phys. Chem. Lett.* **2021**, *12* (32), 7792–7796. https://doi.org/10.1021/acs.jpclett.1c02179.

(23) Zou, Z.; Zhang, Z.; Xu, J.; Yu, Z.; Cheng, M.; Xiong, R.; Lu, Z.; Liu, Y.; Shi, J. Thermochromic, Threshold Switching, and Optical Properties of Cr-Doped VO2 Thin Films. *Journal of Alloys and Compounds* **2019**, *806*, 310–315. https://doi.org/10.1016/j.jallcom.2019.07.264.

(24) Huang, Z. Improvement of Phase Transition Properties of Magnetron Sputtered W-Doped VO2 Films by Post-Annealing Approach. *Journal of Materials Science* **2020**, 11.

(25) Hanlon, T. J.; Coath, J. A.; Richardson, M. A. Molybdenum-Doped Vanadium Dioxide Coatings on Glass Produced by the Aqueous Sol–Gel Method. *Thin Solid Films* **2003**, *436* (2), 269–272. https://doi.org/10.1016/S0040-6090(03)00602-3.

(26) Mlyuka, N. R.; Niklasson, G. A.; Granqvist, C. G. Mg Doping of Thermochromic VO2 Films Enhances the Optical Transmittance and Decreases the Metal-Insulator Transition Temperature. *Appl. Phys. Lett.* **2009**, *95* (17), 171909. https://doi.org/10.1063/1.3229949.

(27) Tan, X.; Yao, T.; Long, R.; Sun, Z.; Feng, Y.; Cheng, H.; Yuan, X.; Zhang, W.; Liu, Q.; Wu, C.; Xie, Y.; Wei, S. Unraveling Metal-Insulator Transition Mechanism of VO2 Triggered by Tungsten Doping. *Sci Rep* **2012**, *2* (1), 466. https://doi.org/10.1038/srep00466.

(28) Koch, D.; Chaker, M. The Origin of the Thermochromic Property Changes in Doped Vanadium Dioxide. *ACS Appl. Mater. Interfaces* **2022**, *14* (20), 23928–23943. https://doi.org/10.1021/acsami.2c02070.

(29) Wu, Y.; Fan, L.; Huang, W.; Chen, S.; Chen, S.; Chen, F.; Zou, C.; Wu, Z. Depressed Transition Temperature of $W_xV_{1-x}O_2$: Mechanistic Insights from the X-Ray Absorption Fine Structure (XAFS) Spectroscopy. *Phys. Chem. Chem. Phys.* **2014**, *16* (33), 17705. https://doi.org/10.1039/C4CP01661K.

(30) Wilson, C. E.; Gibson, A. E.; Cuillier, P. M.; Li, C.-H.; Crosby, P. H. N.; Trigg, E. B.; Najmr, S.; Murray, C. B.; Jinschek, J. R.; Doan-Nguyen, V. Local Structure Elucidation of Tungsten-Substituted Vanadium Dioxide ($V_{1-x}W_xO_2$). *Sci Rep* **2022**, *12* (1), 14767. https://doi.org/10.1038/s41598-022-18575-0.

(31) Azmat, M.; Haibo, J.; Naseem, K.; Ling, C.; Li, J. A Comparative Study Uncovering the Different Effect of Nb, Mo and W Dopants on Phase Transition of Vanadium Dioxide. *Journal of Physics and Chemistry of Solids* **2023**, *180*, 111439. https://doi.org/10.1016/j.jpcs.2023.111439.

(32) Liang, Z.; Zhao, L.; Meng, W.; Zhong, C.; Wei, S.; Dong, B.; Xu, Z.; Wan, L.; Wang, S. Tungsten-Doped Vanadium Dioxide Thin Films as Smart Windows with Self-Cleaning and Energy-Saving Functions. *Journal of Alloys and Compounds* **2017**, *694*, 124–131. https://doi.org/10.1016/j.jallcom.2016.09.315.

(33) Appavoo, K.; Nag, J.; Wang, B.; Luo, W.; Duscher, G.; Payzant, E. A.; Sfeir, M. Y.; Pantelides, S. T.; Haglund, R. F. J. Doping-Driven Electronic and Lattice Dynamics in the Phase-Change Material Vanadium Dioxide. *Physical Review B* **2020**, *102* (11), 10. https://doi.org/10.1103/PhysRevB.102.115148.

(34) Jung, K. H.; Yun, S. J.; Slusar, T.; Kim, H.-T.; Roh, T. M. Highly Transparent Ultrathin Vanadium Dioxide Films with Temperature-Dependent Infrared Reflectance for Smart Windows. *Applied Surface Science* **2022**, *589*, 152962. https://doi.org/10.1016/j.apsusc.2022.152962.

(35) Théry, V.; Boulle, A.; Crunteanu, A.; Orlianges, J. C.; Beaumont, A.; Mayet, R.; Mennai, A.; Cosset, F.; Bessaudou, A.; Fabert, M. Role of Thermal Strain in the Metal-Insulator and Structural Phase Transition of Epitaxial VO2 Films. *Phys. Rev. B* **2016**, *93* (18), 184106. https://doi.org/10.1103/PhysRevB.93.184106.

(36) Gentils, A.; Cabet, C. Investigating Radiation Damage in Nuclear Energy Materials Using JANNuS Multiple Ion Beams. *Nuclear Instruments and Methods in Physics Research Section B: Beam Interactions with Materials and Atoms* **2019**, *447*, 107–112. https://doi.org/10.1016/j.nimb.2019.03.039.

(37) Mayer, M. SIMNRA, a Simulation Program for the Analysis of NRA, RBS and ERDA. In *AIP Conference Proceedings*; AIP: Denton, Texas (USA), 1999; pp 541–544. https://doi.org/10.1063/1.59188.

(38) Wertheim, G. K.; Butler, M. A.; West, K. W.; Buchanan, D. N. E. Determination of the Gaussian and Lorentzian Content of Experimental Line Shapes. *Review of Scientific Instruments* **1974**, *45* (11), 1369–1371. https://doi.org/10.1063/1.1686503.

(39) Rahm, M.; Li, J.-S.; Padilla, W. J. THz Wave Modulators: A Brief Review on Different Modulation Techniques. *J Infrared Milli Terahz Waves* **2013**, *34* (1), 1–27. https://doi.org/10.1007/s10762-012-9946-2.

(40) Rajeswaran, B.; Umarji, A. M. Effect of W Addition on the Electrical Switching of VO2 Thin Films. *AIP Advances* **2016**, *6* (3), 035215. https://doi.org/10.1063/1.4944855.

(41) Sun, T. Classical Size Effect In Copper Thin Films: Impact Of Surface And Grain Boundary Scattering On Resistivity. *Electronic Theses and Dissertations.* **2009**, *3927*.

(42) Bakonyi, I.; Isnaini, V. A.; Kolonits, T.; Czigány, Zs.; Gubicza, J.; Varga, L. K.; Tóth-Kádár, E.; Pogány, L.; Péter, L.; Ebert, H. The Specific Grain-Boundary Electrical Resistivity of Ni. *Philosophical Magazine* **2019**, *99* (9), 1139–1162. https://doi.org/10.1080/14786435.2019.1580399.

(43) Narayan, J.; Bhosle, V. M. Phase Transition and Critical Issues in Structure-Property Correlations of Vanadium Oxide. *Journal of Applied Physics* **2006**, *100* (10), 103524. https://doi.org/10.1063/1.2384798.

(44) Karaoglan-Bebek, G.; Hoque, M. N. F.; Holtz, M.; Fan, Z.; Bernussi, A. A. Continuous Tuning of W-Doped $VO_2$ Optical Properties for Terahertz Analog Applications. *Appl. Phys. Lett.* **2014**, *105* (20), 201902. https://doi.org/10.1063/1.4902056.

(45) Ivanov, A. V.; Tatarenko, A. Yu.; Gorodetsky, A. A.; Makarevich, O. N.; Navarro-Cía, M.; Makarevich, A. M.; Kaul, A. R.; Eliseev, A. A.; Boytsova, O. V. Fabrication of Epitaxial W-Doped $VO_2$ Nanostructured Films for Terahertz Modulation Using the Solvothermal Process. *ACS Appl. Nano Mater.* **2021**, *4* (10), 10592–10600. https://doi.org/10.1021/acsanm.1c02081.

(46) Zhao, Y.; Chen, C.; Pan, X.; Zhu, Y.; Holtz, M.; Bernussi, A.; Fan, Z. Tuning the Properties of $VO_2$ Thin Films through Growth Temperature for Infrared and Terahertz Modulation Applications. *Journal of Applied Physics* **2013**, *114* (11), 113509. https://doi.org/10.1063/1.4821846.

(47) Koughia, C.; Gunes, O.; Zhang, C.; Wen, S.-J.; Wong, R.; Yang, Q.; Kasap, S. O. Topology of Conductive Clusters in Sputtered High-Quality $VO_2$ Thin Films on the Brink of Percolation Threshold during Insulator-to-Metal and Metal-to-Insulator Transitions. *Journal of Vacuum Science & Technology A* **2020**, *38* (6), 063401. https://doi.org/10.1116/6.0000443.

(48) Okimura, K.; Ezreena, N.; Sasakawa, Y.; Sakai, J. Electric-Field-Induced Multistep Resistance Switching in Planar $VO_2$ / *c*-$Al_2O_3$ Structure. *Jpn. J. Appl. Phys.* **2009**, *48* (6), 065003. https://doi.org/10.1143/JJAP.48.065003.

(49) Cheng, S.; Lee, M.-H.; Li, X.; Fratino, L.; Tesler, F.; Han, M.-G.; del Valle, J.; Dynes, R. C.; Rozenberg, M. J.; Schuller, I. K.; Zhu, Y. *Operando* Characterization of Conductive Filaments during Resistive Switching in Mott $VO_2$. *Proc Natl Acad Sci USA* **2021**, *118* (9), e2013676118. https://doi.org/10.1073/pnas.2013676118.

(50) Gu, M. J.; Lin, S.; Xu, X. F.; Wang, C. R.; Wu, B. H.; Cao, J. C. Random-Resistor Network Modeling of Resistance Hysteresis of Vanadium Dioxide Thin Films. *Journal of Applied Physics* **2022**, *132* (1), 015301. https://doi.org/10.1063/5.0093242.

(51) Qazilbash, M. M.; Brehm, M.; Chae, B.-G.; Ho, P.-C.; Andreev, G. O.; Kim, B.-J.; Yun, S. J.; Balatsky, A. V.; Maple, M. B.; Keilmann, F.; Kim, H.-T.; Basov, D. N. Mott Transition in $VO_2$ Revealed by Infrared Spectroscopy and Nano-Imaging. *Science* **2007**, *318* (5857), 1750–1753. https://doi.org/10.1126/science.1150124.

(52) Gopalakrishnan, G.; Ruzmetov, D.; Ramanathan, S. On the Triggering Mechanism for the Metal–Insulator Transition in Thin Film $VO_2$ Devices: Electric Field versus Thermal Effects. *J Mater Sci* **2009**, *44* (19), 5345–5353. https://doi.org/10.1007/s10853-009-3442-7.

(53) Liang, G.; Hu, X.; Yu, X.; Shen, Y.; Li, L. H.; Davies, A. G.; Linfield, E. H.; Liang, H. K.; Zhang, Y.; Yu, S. F.; Wang, Q. J. Integrated Terahertz Graphene Modulator with 100% Modulation Depth. *ACS Photonics* **2015**, *2* (11), 1559–1566. https://doi.org/10.1021/acsphotonics.5b00317.

(54) Fan, Z.; Geng, Z.; Lv, X.; Su, Y.; Yang, Y.; Liu, J.; Chen, H. Optical Controlled Terahertz Modulator Based on Tungsten Disulfide Nanosheet. *Sci Rep* **2017**, *7* (1), 14828. https://doi.org/10.1038/s41598-017-13864-5.

(55) Jiang, M.; Xu, X.; Hu, F.; Du, H.; Zhang, L.; Zou, Y. Low-Voltage Triggered $VO_2$ Hybrid Metasurface Used for Amplitude Modulation of Terahertz Orthogonal Modes. *J. Lightwave Technol.* **2022**, *40* (1), 156–162. https://doi.org/10.1109/JLT.2021.3120730.

(56) Han, C.; Parrott, E. P. J.; Humbert, G.; Crunteanu, A.; Pickwell-MacPherson, E. Broadband Modulation of Terahertz Waves through Electrically Driven Hybrid Bowtie Antenna-VO2 Devices. *Sci Rep* **2017**, *7* (1), 12725. https://doi.org/10.1038/s41598-017-13085-w.

(57) Liang, W.; Jiang, Y.; Guo, J.; Li, N.; Qiu, W.; Yang, H.; Ji, Y.; Luo, S. Van Der Waals Heteroepitaxial $VO_2$ /Mica Films with Extremely Low Optical Trigger Threshold and Large THz Field Modulation Depth. *Advanced Optical Materials* **2019**, *7* (20), 1900647. https://doi.org/10.1002/adom.201900647.

# Supporting information

# Electrically activated W-doped VO$_2$ films for reliable, large-area, broadband THz waves modulators

*Eduard-Nicolae Sirjita[a,b]\*, Alexandre Boulle[b]\*, Jean-Christophe Orlianges[a], Richard Mayet[b], Aurélien Debelle[c], Lionel Thomé[c], Maggy Colas[b], Julie Cornette[b], Aurelian Crunteanu[a]\**

[a] XLIM Research Institute, CNRS/ University of Limoges, France;

[b] Institut de Recherche sur les Céramiques (IRCER), CNRS UMR 7315, University of Limoges France

[c] IJCLab, Université Paris-Saclay, CNRS/IN2P3, Orsay, France.

\*Corresponding authors:

eduard.sirjita@xlim.fr, alexandre.boulle@cnrs.fr, aurelian.crunteanu@xlim.fr

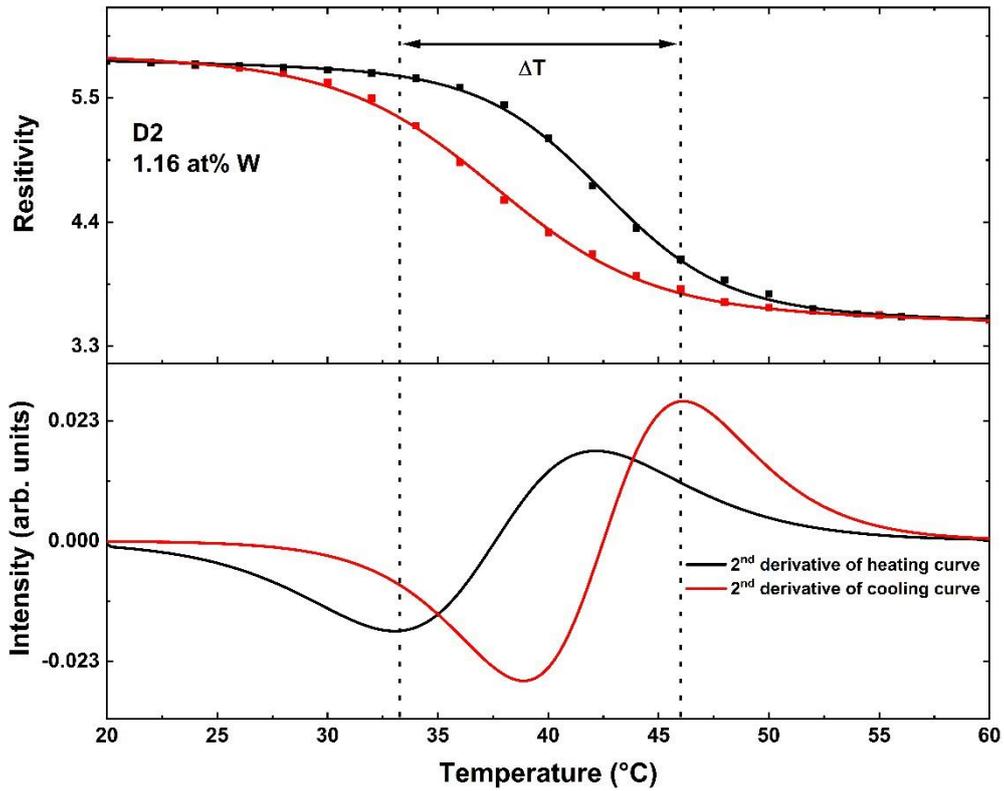

**Figure S1.** A typical example of a temperature-dependent resistivity measurement (sample D2). In the upper panel the points are the measurements, whereas the continuous line is a fit with equation 2 in the article. $T_{IMT}$ and $T_{MIT}$ are taken directly from the b parameter of the fitting function presented in equation 2 and correspond to the inflection points of the lines. The lower panel shows the transition width as measured from the second derivative.

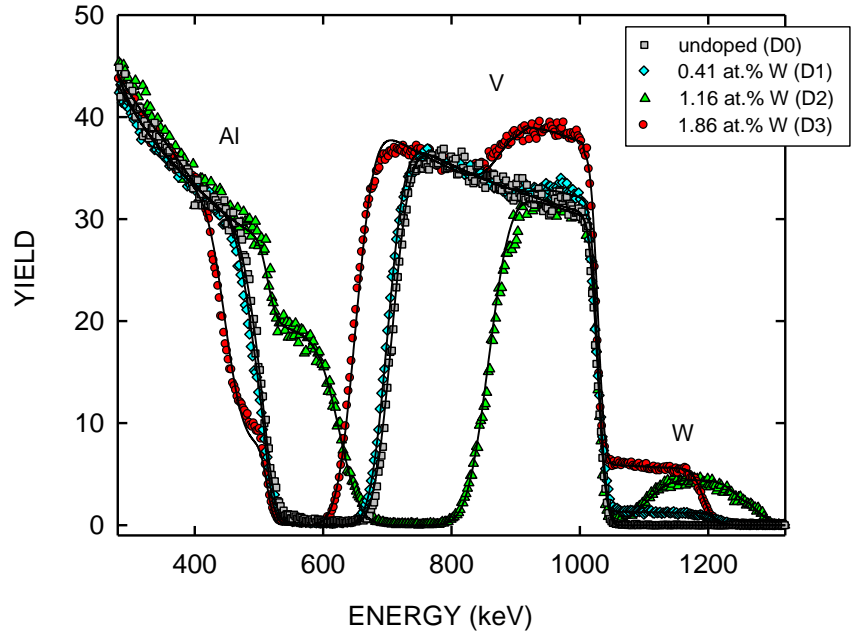

**Figure S2.** The Rutherfordback scattering spectrometry (RBS) spectra registered in a random orientation on thin $VO_2$ layers deposited on a sapphire substrate and doped with several W concentrations. The spectra exhibit several components: a signal at low energy (below ~600 keV) arising from the $Al_2O_3$ substrate; a plateau at middle energy (between ~600 and ~1050 keV) due to the $VO_2$ films; a signal at high energy (above ~1050 keV) coming from the W doping. The solid lines in Figure S1 are simulations of RBS spectra with the SIMNRA code [1,2] that allow providing: (i) the thickness and stoichiometry of the $VO_2$ layers; (ii) the concentration profiles of the W dopants.

Sample D0 was deposited without any W doping so that the RBS spectrum was fitted using only a single layer of $VO_{1.9}$ that implies a slight lack of oxygen. For D1-D3 samples, W dopant concentrations profiles were added to the simulations. Thus, sample D1 which was expected to be the least doped film, reveals a corresponds to 0.41 at% of W. Sample D2 and D3 correspond to 1.16 and 1.86 at% of W.

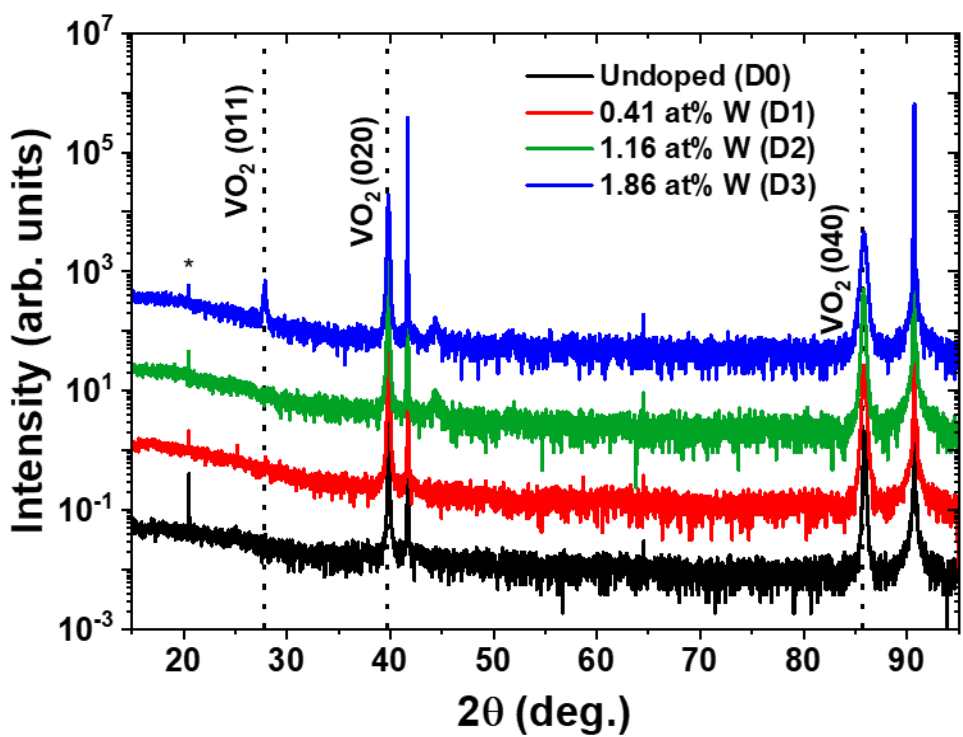

**Figure S3.** θ-2θ scans of the films D0-D3. All the XRD scans present an intense peak at around 39° and 85° which correspond to the 020 and 040 reflections from $VO_2$, in agreement with the (010) orientation mentioned in the article. The 006 and 0012 peaks of sapphire are observed at 41.7° and 90.7°. For the highest level of doping a weak peak around ~30° that can be attributed to the (011) reflection of $VO_2$ can be observed. The origin of the presence of this weak secondary orientation is unclear at the moment, but it was not found to have any sort of influence on the results presented here. The very weak intensity of this peak points to a low volume of the associated orientation, so that it can be easily neglected. The peaks labelled with "*" correspond to forbidden reflections of sapphire which appear due to multiple diffraction in the high quality $Al_2O_3$ substrates.

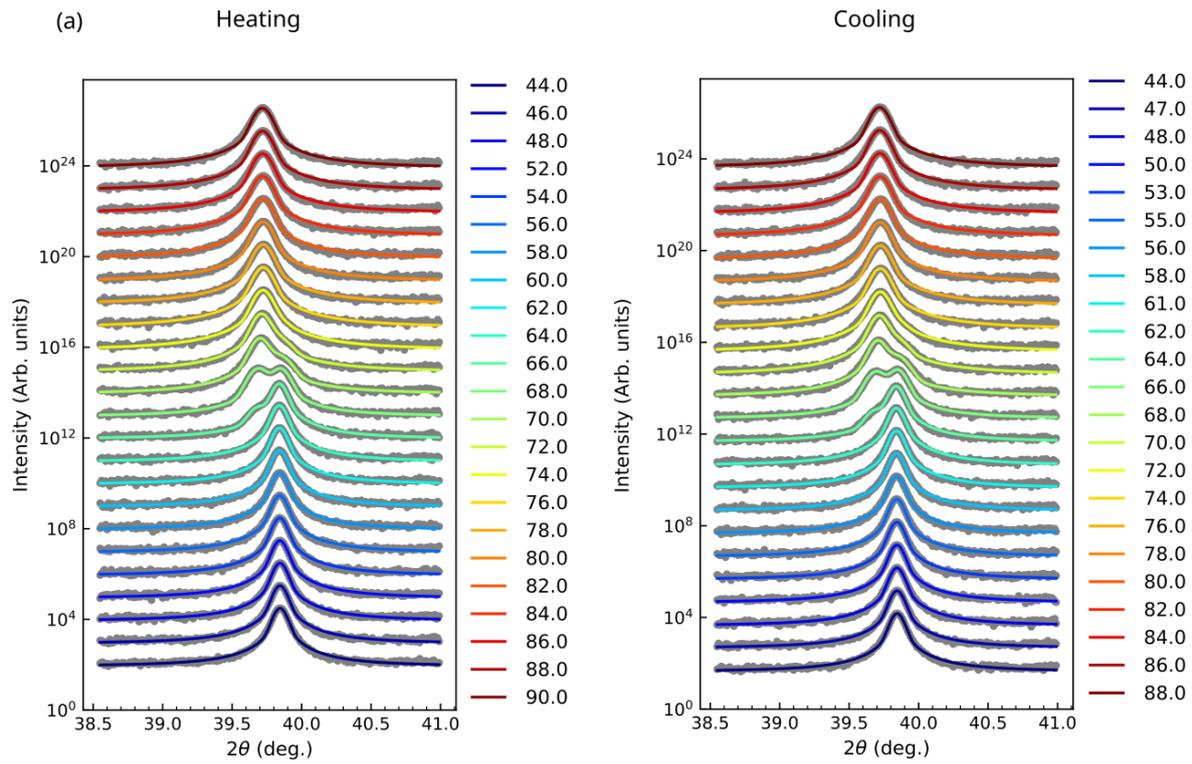

(a)

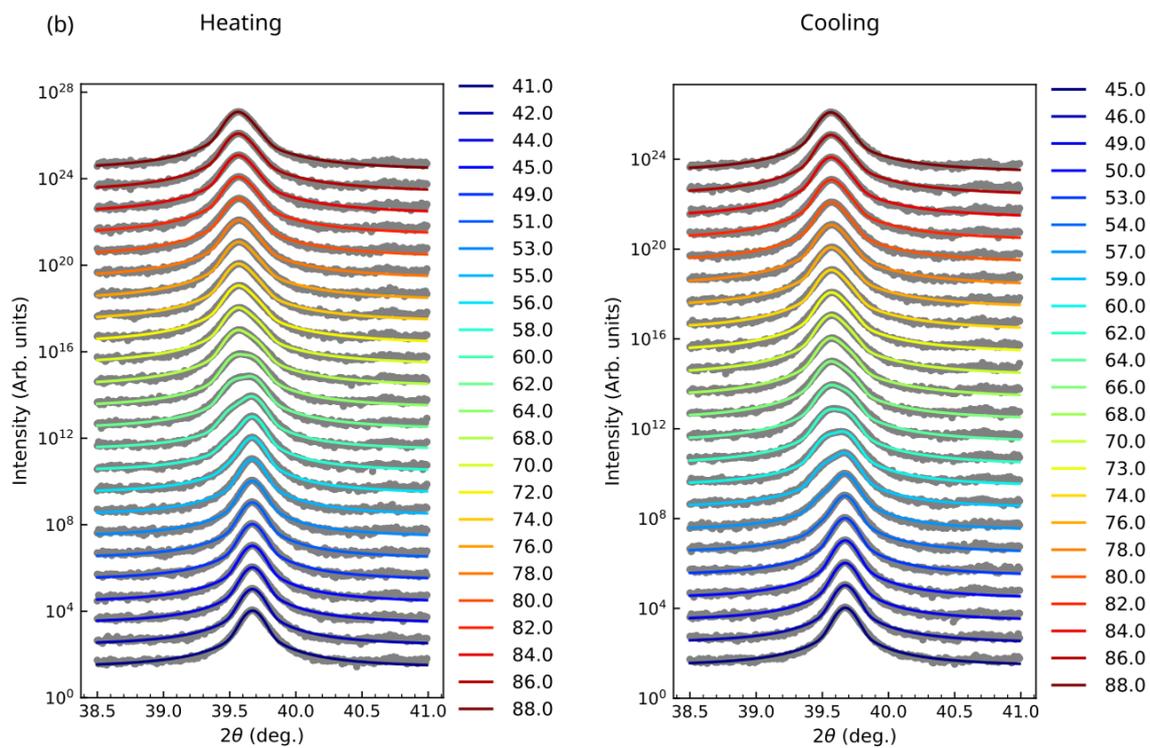

(b)

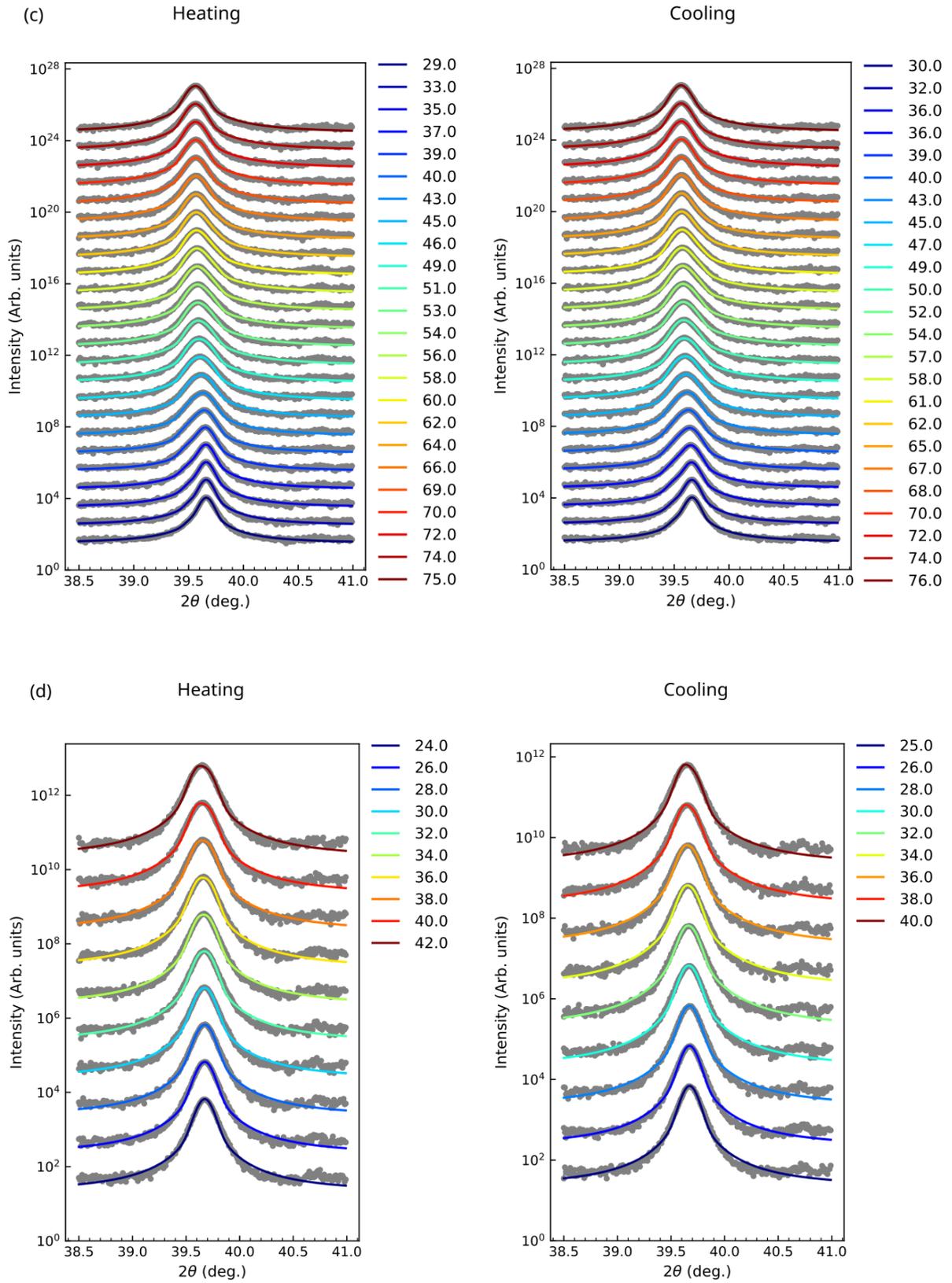

**Figure S4**. Fit of the XRD temperature-dependent data with a model based on two pseudo-Voigt functions (one for the M1 and another for the R phase) and a linear background. (a), (b),

(c), (d): D0, D1, D2, D3, respectively. The intensity is plotted on log scale and the curves are shifted vertically for clarity.

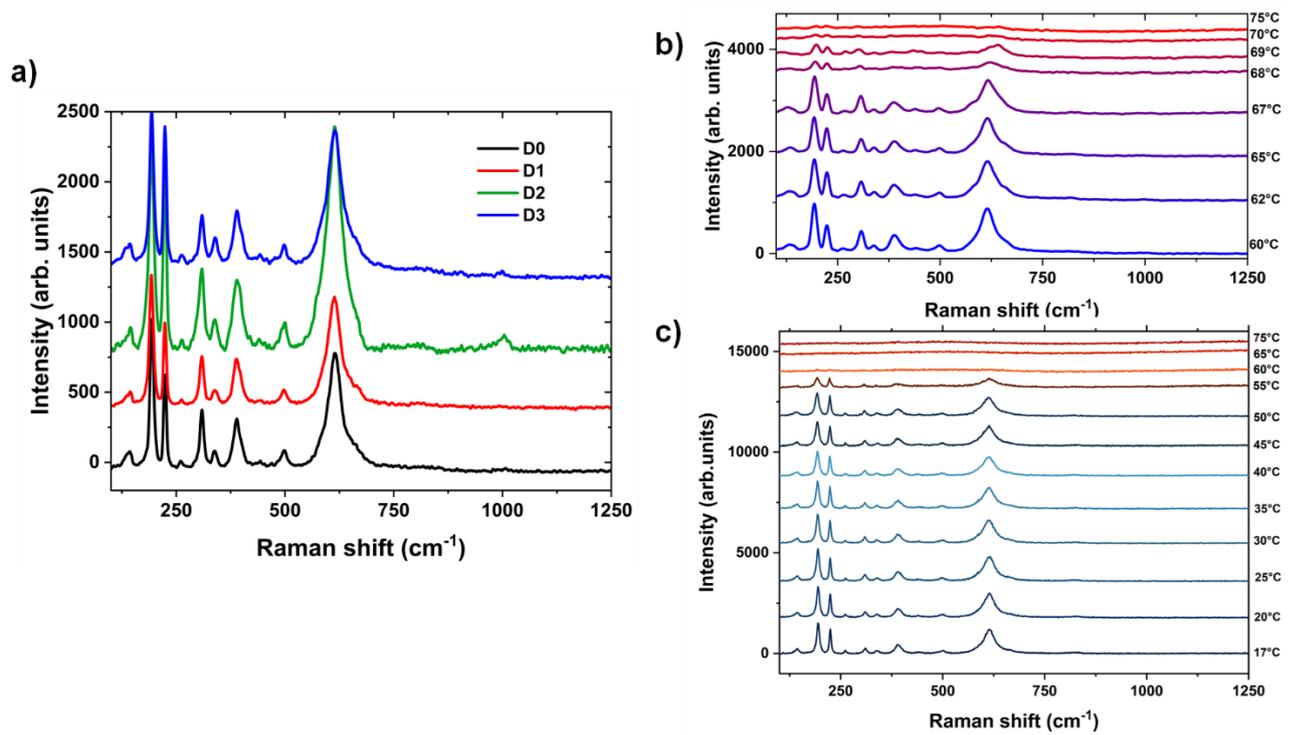

**Figure S5.** The Raman spectra of all films, recorded at room temperature are shown in (a). The temperature-dependent Raman spectra sample D0 and D3 are shown in (b) and (c), respectively. The room temperature spectra have very similar features of the spectra, with no obvious remarkable shifts between the films. All films present typical M1 phase $VO_2$ peaks at ~144, 193, 223, 259, 310, 339, 387, 442, 499, 613 cm$^{-1}$ [3].

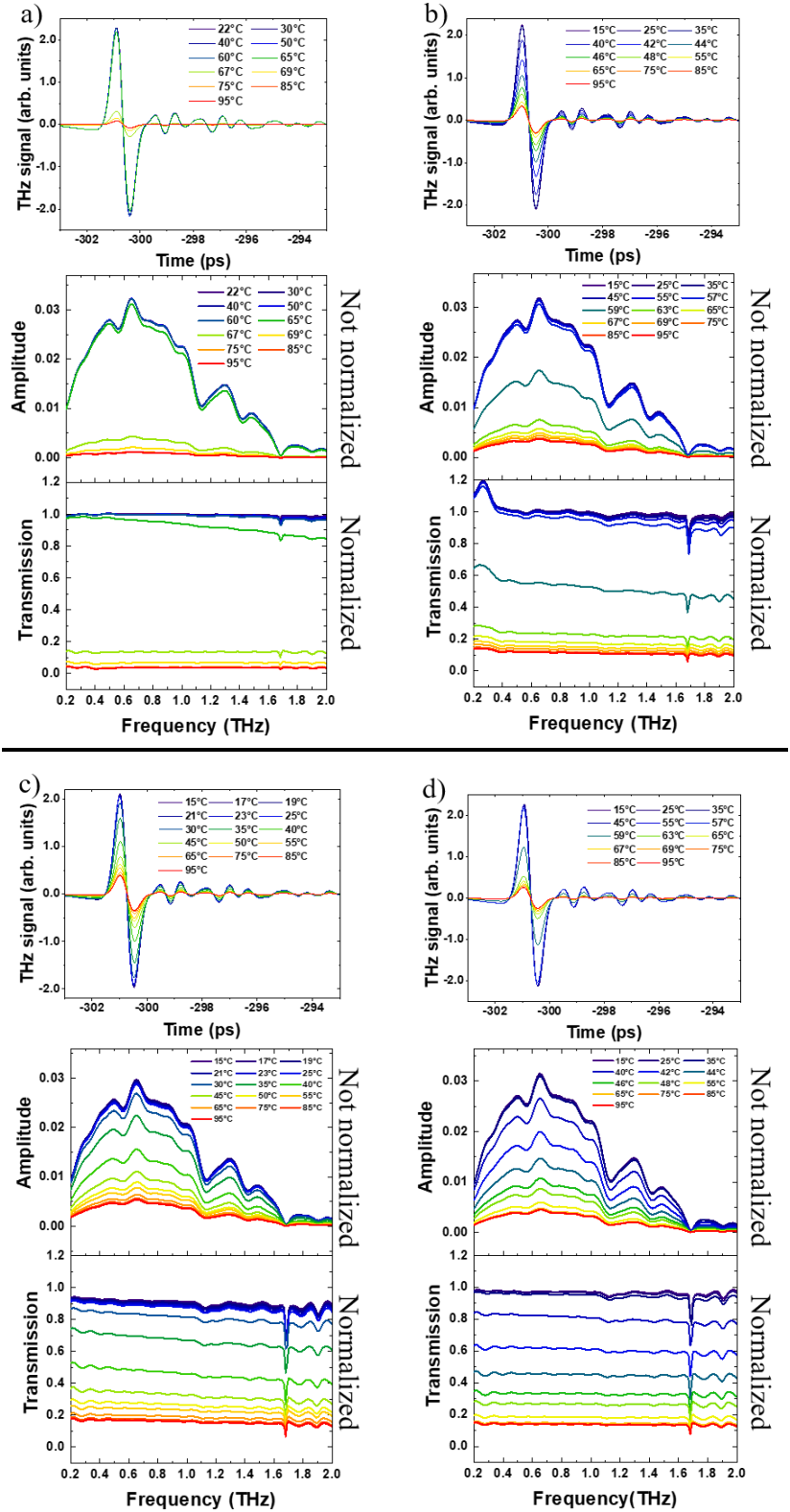

**Figure S6.** The THz characterizations of samples D0-D4 are shown in (a-d), respectively. In each figure, the temperature variation of THz time-domain spectra is given in the upper panel, the frequency domain spectra (FFT transformation of the time-domain ones) in the middle

panel, and normalized frequency domain spectra (sapphire substrate + ambient atmosphere influence removed) in the lower panel.

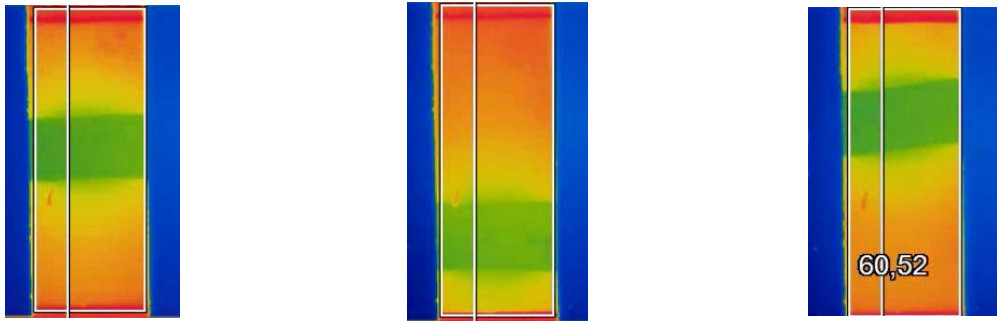

**Figure S7.** IR images illustrating the random metallic channel positions at the onset of the electrically induced insulator-to-metal transition in a D0 type device.

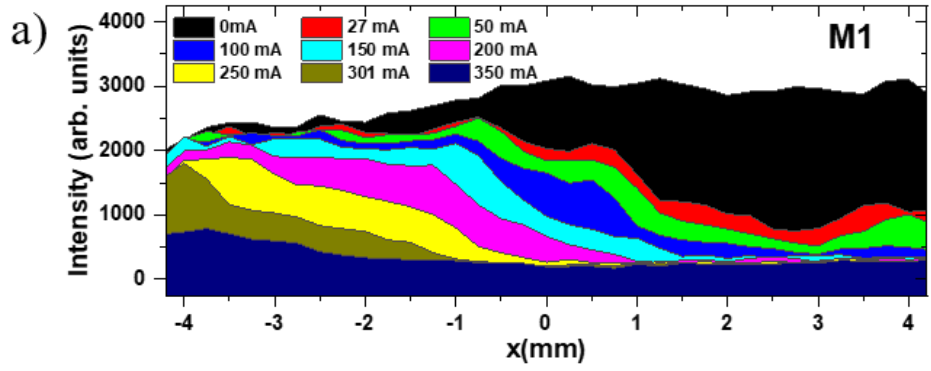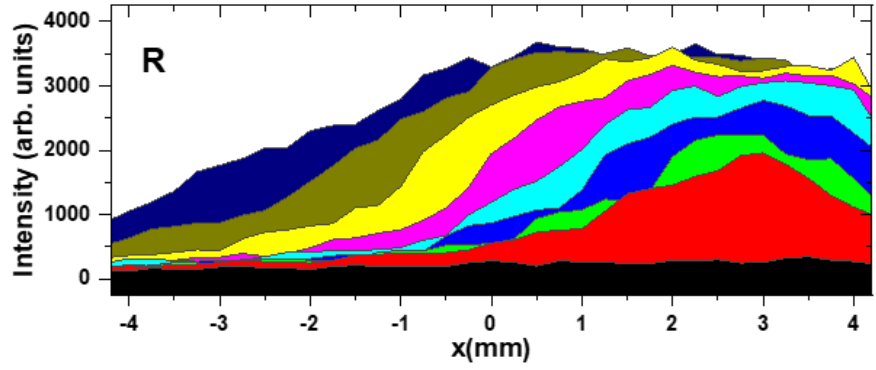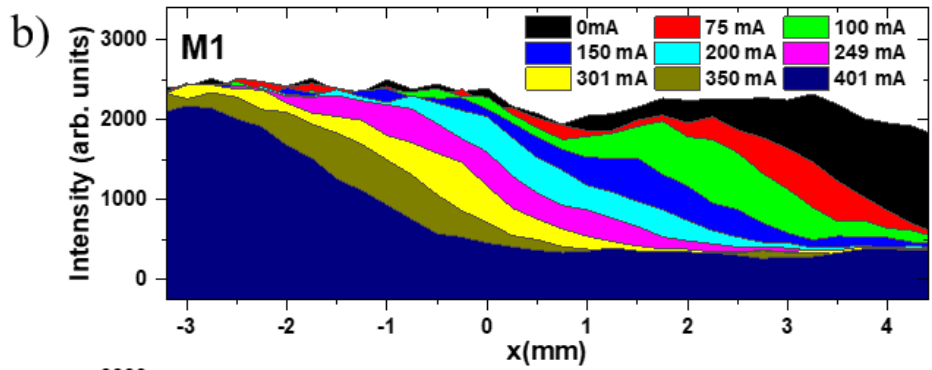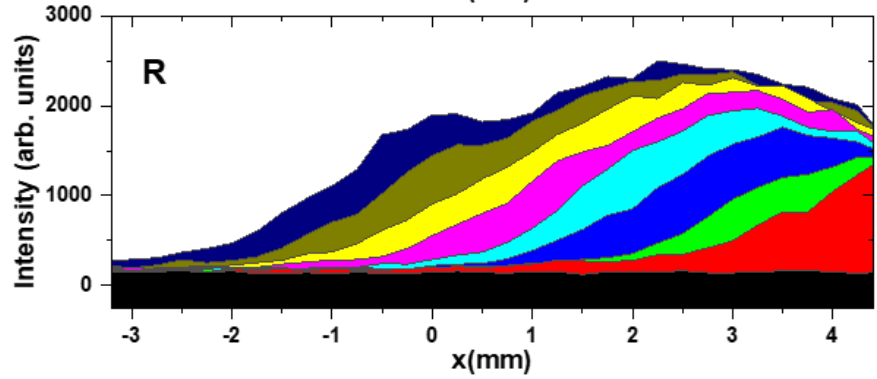

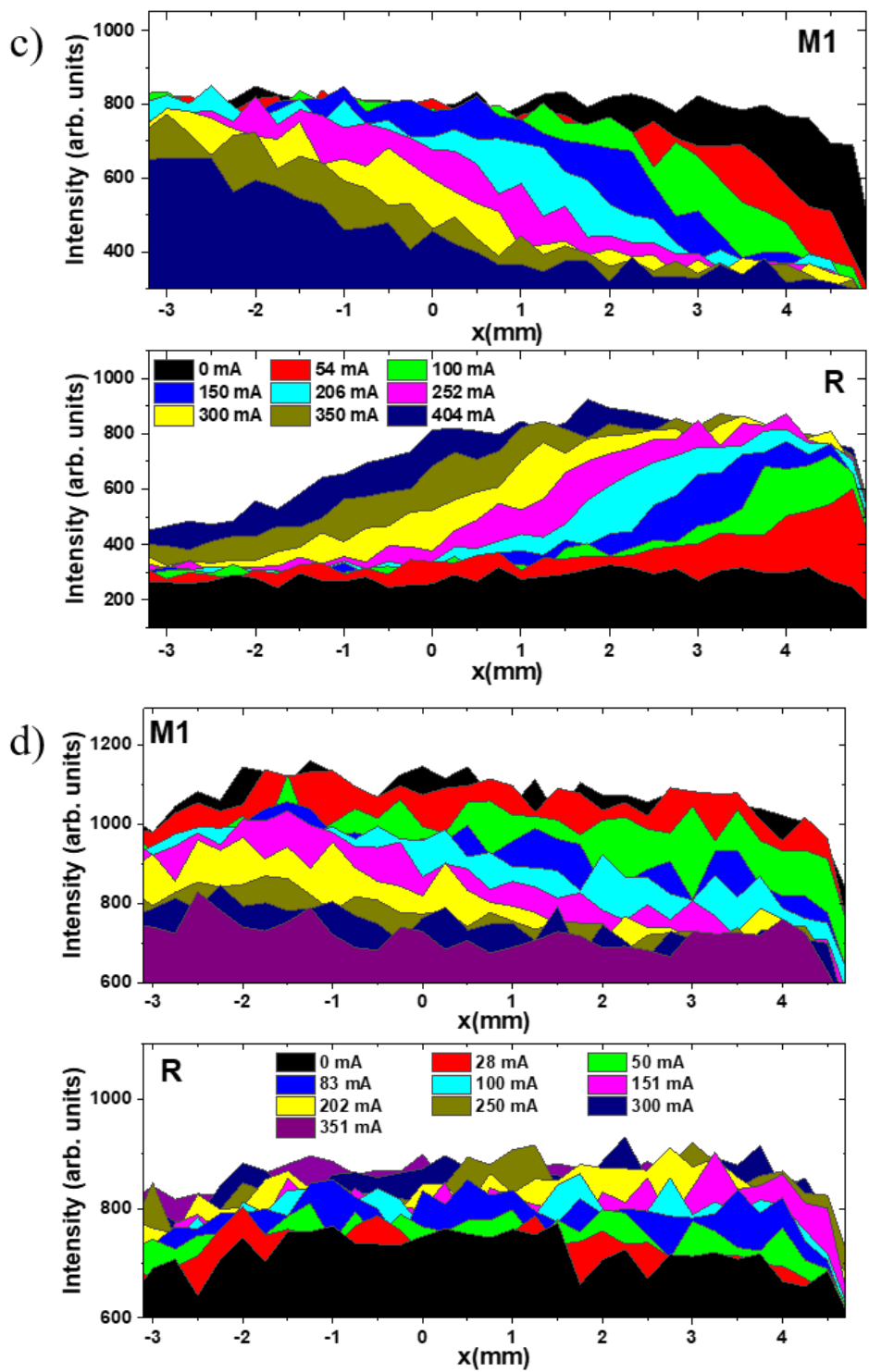

**Figure S8.** XRD scans along the VO$_2$ –based devices for showing the complementarity of the M1 and R intensity evolution for increasing applied currents across the a) D0, b) D1, c) D2 and d) D3 samples.

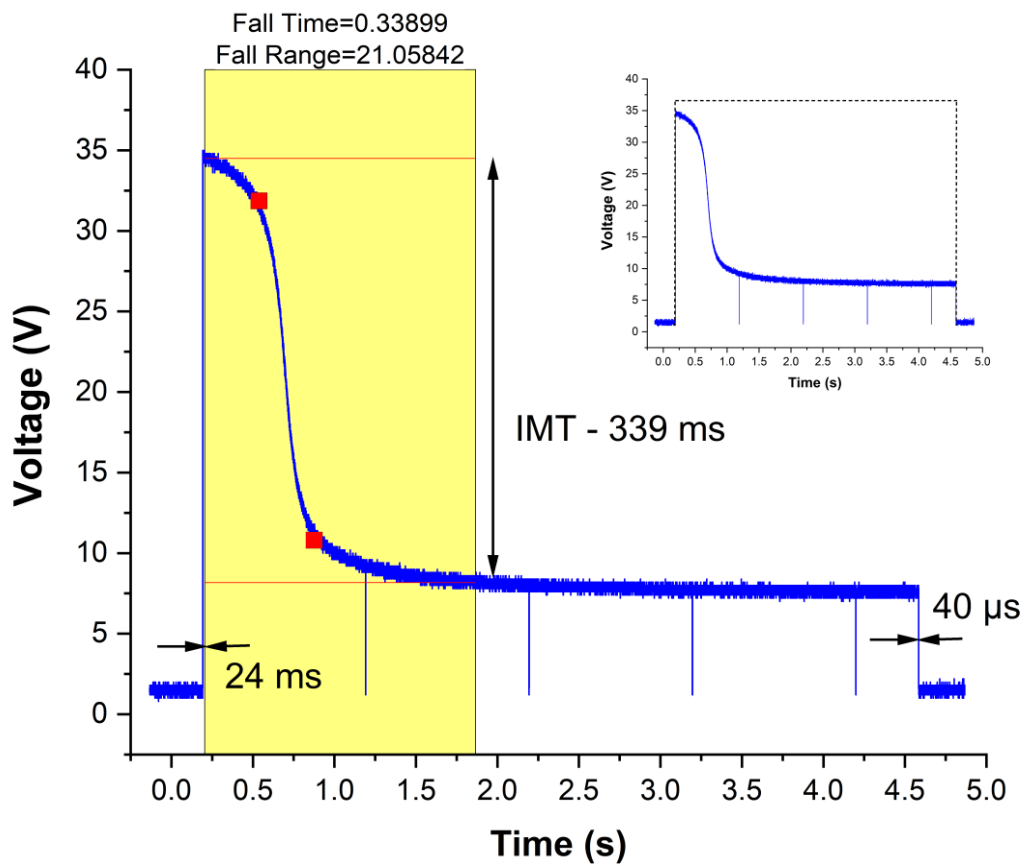

**Figure S9.** Voltage response across the D3 device (oscilloscope recorded blue curve) when applying a square-shaped current pulse (400 mA amplitude, length 4.4 s- black dotted curve in the inset). Upon the application of the pulse, the device experiences a fast voltage drop following its high-to-low resistance change indicating that the W-doped $VO_2$ material was performing an insulator –to-metal transition (IMT).

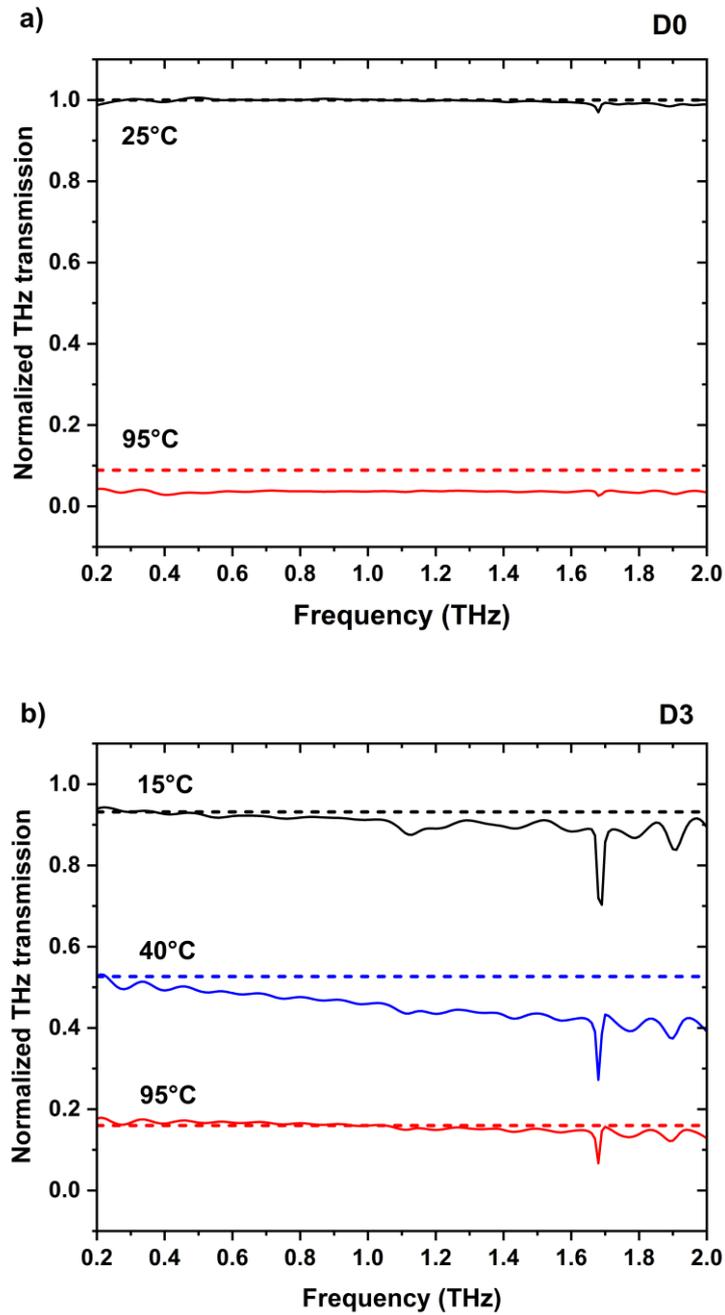

**Figure S10.** Measured (solid lines) and simulated (dotted lines) normalized THz transmissions at different temperatures (hence different corresponding electrical resistivities/ conductivities as represented in **Figure 2 b** in the manuscript) for the devices a) D0 -pure $VO_2$ layer) and b) D3-W-doped $VO_2$ layer.

The complex frequency- dependent transmitted electrical fields in **Figures 3 a.** and **b. and Figure S6** above are related to the complex conductivity of the films by the equation 1 below [5-7]:

$$\frac{\check{T}_{film+substrate}(\omega)}{\check{T}_{substrate}(\omega)} = \frac{1+n_{substrate}}{1+n_{substrate}+Z_0\check{\sigma}(\omega)d} \qquad (1)$$

where $Z_0$ is the impedance of the free space and $n_{substrate} \approx 3$ is the nearly frequency-independent index of refraction of the sapphire substrate and $d$ is the thin film thickness. The solid lines in **Figure S10 a) and b)** are the measured normalized THz transmissions ($T_{film+substrate}(\omega)/T_{substrate}(\omega)$) through the D0 and D3 devices, at different temperatures. The dotted lines are the simulated theoretical curves of the relative terahertz transmission of the films calculated from Eq. 1 above. For each temperature, we assumed a real and frequency-independent conductivity, $\sigma(\omega) \approx \sigma_{DC}$ which was extracted from the electrical resistivity variation with temperature in **Figure 2b** in the manuscript. A good agreement between the theoretical simulation and the experiment results is evident from the curves in Figure S10, especially at low terahertz frequencies.

References

[1]  M. Mayer, In *AIP Conference Proceedings*, AIP, Denton, Texas (USA), **1999**, pp. 541–544.

[2]  A. Gentils, C. Cabet, *Nuclear Instruments and Methods in Physics Research Section B: Beam Interactions with Materials and Atoms* **2019**, *447*, 107.

[3]  P. Schilbe, *Physica B: Condensed Matter* **2002**, *316–317*, 600.

[4]  K. Okimura, N. Hanis Azhan, T. Hajiri, S. Kimura, M. Zaghrioui, J. Sakai, *Journal of Applied Physics* **2014**, *115*, 153501.

[5]  M. Walther, D. G. Cooke, C. Sherstan, M. Hajar, M. R. Freeman, and F. A. Hegmann, "Terahertz conductivity of thin gold films at the metal-insulator percolation transition", *Physical Review B* **2007**, 76, 125408, DOI: 10.1103/PhysRevB.76.125408

[6]  G. Ma; D. Li; H. Ma; J. Shen; C. Wu; J. Ge; S. Hu; N. Dai, "Carrier concentration dependence of terahertz transmission on conducting ZnO films", *Appl. Phys. Lett.* **2008**, 93, 211101, https://doi.org/10.1063/1.3036708

[7]  Karaoglan-Bebek, G.; Hoque, M. N. F.; Holtz, M.; Fan, Z.; Bernussi, A. A. Continuous Tuning of W-Doped $VO_2$ Optical Properties for Terahertz Analog Applications. *Appl. Phys. Lett.* **2014**, *105* (20), 201902. https://doi.org/10.1063/1.4902056.